\documentclass{new_cls}

\usepackage{mathtools}

\usepackage[dvipsnames]{xcolor}

\let\Re\relax
\let\Im\relax
\DeclareMathOperator{\Re}{\mathrm{Re}}
\DeclareMathOperator{\Im}{\mathrm{Im}}

\newcommand{\pp}[2]{\frac{\partial #1}{\partial #2}}
\newcommand{\DD}[2]{\frac{\mathrm{D} #1}{\mathrm{D} #2}}
\newcommand{\mean}[1]{\overline{#1}}

\title{Energetics of frontogenesis in \\ simple balanced models}

\author{Yue Bai}
\author{J\"orn Callies}
\affil{California Institute of Technology, Pasadena, California}
\corr{Yue Bai, \href{mailto:ybbai@caltech.edu}{ybbai@caltech.edu} \\[\bigskipamount] This work has been accepted to the Journal of Physical Oceanography. The American Meteorological Society does not guarantee that the copy provided here is an accurate copy of the version of record.}

\begin{document}

\nolinenumbers

\section*{Abstract}

Submesoscale fronts have been proposed to act as conduits funneling kinetic energy from geostrophically constrained mesoscale eddies down to small scales, where the energy can be dissipated.
Realistic primitive-equation simulations suggest that the downscale energy transfer at submesoscales is strongly concentrated in sharpening fronts, with convergence on the cyclonic side playing a dominant role.
This study explores how much of this phenomenology can be captured by simple quasi-geostrophic (QG) and semi-geostrophic (SG) theories of frontogenesis.
An analysis of the kinetic energy budget of classical strain-induced frontogenesis shows that there is downscale transfer in both QG and SG fronts.
In QG frontogenesis, the narrowing jet along the sharpening front is powered by buoyancy production, but geostrophic downscale transfer due to the strain field plays an important role.
In SG frontogenesis, where frontal sharpening is accelerated by the ageostrophic cross-frontal circulation, ageostrophic advection enhances this downscale transfer on the cyclonic side of the front, where the surface flow is convergent.
On the anticyclonic side, where the ageostrophic circulation is divergent, the ageostrophic scale transfer of kinetic energy is upscale and partially offsets the geostrophic downscale transfer.
While matching patterns in realistic primitive-equation simulations, the ageostrophic scale transfers remain important but not dominant in the adiabatic and frictionless theory considered here.
This suggests that processes absent from these theories may play an important role.
For example, mixed-layer turbulence may enhance the ageostrophic circulation and the associated dipole of scale transfers.
The advection of ageostrophic momentum may also be important in some circumstances.

\section{Introduction}

Kinetic energy in the ocean is distributed across a wide range of spatial scales, with mesoscale eddies containing some 80\% of the total \parencite{ferrari2009,chelton2011}.
At mesoscales, the ocean is strongly influenced by Earth's rotation and stratification, placing mesoscale eddies in the quasi-geostrophic regime.
In this regime, the classical dual-cascade paradigm of geostrophic turbulence is thought to apply \parencite{rhines_dynamics_1977,salmon1978}. Large-scale available potential energy is transferred downscale and converted to kinetic energy by mesoscale baroclinic instability.
Due to the geostrophic constraint, nonlinear triad interactions subsequently transfer this kinetic energy back to larger scales \parencite{kraichnan1967,charney1971,scott2005,tulloch2011,arbic_eddy_2013}.
This inverse cascade prompts the question of how kinetic energy is dissipated to balance the mesoscale source---somehow, the energy must be transferred to small scales where viscous dissipation can act.
Many processes are thought to play a role. Interaction with the bottom, especially at western boundaries \parencite{zhai_significant_2010}, can generate boundary layer turbulence and produce dissipation \parencite[e.g.,][]{ferrari2009}. Flow over topography also produces lee waves, whose breaking allows them exert a net drag \parencite[e.g.,][]{nikurashin_routes_2013}. Wind forcing, although acting as a net source of kinetic energy at large scales, produces a top drag at mesoscales through the current dependence of the wind stress \parencite{dewar_effects_1987,duhaut2006,renault2016}, and it can cause symmetric instabilities that extract kinetic energy from the geostrophic flow \parencite{thomas_symmetric_2013}. The interaction with internal waves, especially wind-forced near-inertial waves, can extract energy from mesoscale eddies \parencite{xie_generalised-lagrangian-mean_2015,rocha_stimulated_2018,asselin_penetration_2020}.

Another route to dissipation is the possibility that the geostrophic constraint is broken in the upper ocean and energy is transferred to small scales by energetic submesoscale turbulence \parencite{capet2008-3,muller_routes_2011}. The close analogy between geostrophic turbulence \parencite{charney1971} and two-dimensional turbulence \parencite{kraichnan1967} is broken in the upper ocean, and observations deviate from the geostrophic turbulence prediction of weak submesoscales \parencite{shcherbina2013,callies2015,buckingham16}. The mere presence of the ocean's surface allows for the sharpening of buoyancy gradients and energization of submesoscale turbulence through surface quasi-geostrophic dynamics \parencite{blumen_uniform_1978,held_surface_1995,lapeyre2006,capet_surface_2008}. Furthermore, the discontinuity of potential vorticity (PV) at the base of the mixed layer allows for a submesoscale baroclinic instability of lateral buoyancy gradients \parencite{haine_gravitational_1998,boccaletti2007,fox-kemper_parameterization_1_2008,callies2016}. Especially in deep winter mixed layers, these mixed-layer instabilities appear to inject kinetic energy at submesoscales \parencite{mensa_seasonality_2013,callies2015}, and triad interactions spread it across a wide range of scales \parencite{sasaki2014,callies2016,schubert_submesoscale_2020,lawrence2022}. While both of these scenarios produce a submesoscale kinetic energy transfer that is primarily to large scales, primitive-equation simulations exhibit a small but robust downscale transfer if the resolution is sufficiently high \parencite{capet2008-3}.

\begin{figure}[t]
  \centering
  \includegraphics[width=1\textwidth]{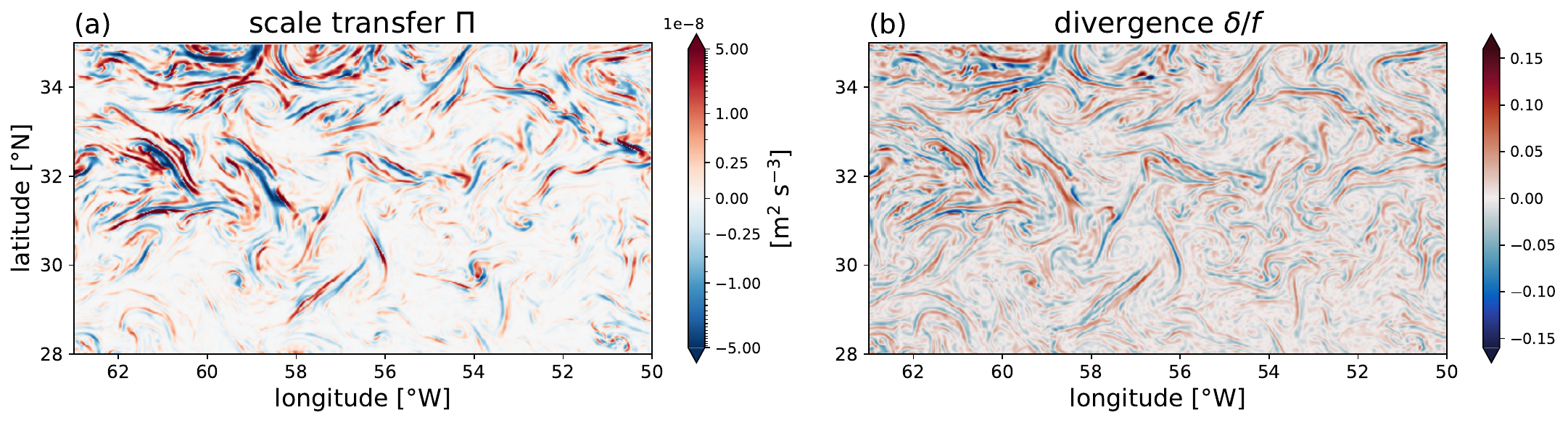}
  \caption{Scale transfer and horizontal divergence at submesoscale fronts in a realistic primitive-equation simulation. Shown are (a)~the scale transfers at a Gaussian filter scale of $\sigma=\qty{3}{\kilo\meter}$ and (b)~a coincident snapshot of divergence filtered to the same scale, exhibiting a tight correspondence (anti-correlation) between the two. Note the arsinh-scaled color map in (a). The simulation is described in \textcite{sinha2023}; we here show surface fields from January~3.}
  \label{fig:non}
\end{figure}

Recently, \textcite{srinivasan2023} linked this downscale transfer of kinetic energy to the formation of fronts. Both flavors of submesoscale turbulence mentioned above tend to generate fronts, either by straining between mesoscale eddies or by lateral shearing induced by mixed-layer instabilities. At the surface, these fronts can sharpen without bound \parencite{stone1966} and collapse into discontinuities \parencite{hoskins72}. When \textcite{srinivasan2023} mapped out the submesoscale scale transfer of kinetic energy, fronts were revealed to play the dominant role (cf., Fig.~\ref{fig:non}). Fronts tended to show a dipole of energy transfer, with a downscale transfer on the cyclonic side of the front, at which the horizontal flow is convergent, dominating over an upscale transfer on the anticyclonic side, at which the horizontal flow is divergent. \textcite{barkan19,srinivasan2023} emphasized the role of convergence in frontogenesis and the downscale energy transfer, and they appealed to mixed-layer turbulence as an important driver of an ageostrophic cross-frontal circulation, convergence, and frontal sharpening.

Nevertheless, the mechanism of a downscale transfer of kinetic energy at sharpening fronts should be generic, and we explore it here using simple balanced models of frontogenesis. We ask how much of the phenomenology of kinetic energy transfer observed in realistic primitive-equation simulations and drifter observations \parencite{barkan19,srinivasan2023} can be reproduced by these simple models. In particular, we consider the strain-induced sharpening of a surface front as the archetype of frontogenesis. We consider the QG dynamics as described by \textcite{stone1966} and the SG dynamics as described by \textcite{hoskins72}. In QG dynamics, as the strain field pushes together a pre-existing buoyancy gradient, a secondary circulation develops that upwells dense water on the light side of the front and downwells light water on the dense side, counteracting its sharpening. At the surface, however, no such vertical advection is present, and the strain field is left to sharpen the buoyancy field exponentially. In SG dynamics, this sharpening at the surface is further accelerated by the horizontal advection by the ageostrophic circulation, which is neglected in QG dynamics but taken into account in SG dynamics. The ageostrophic circulation pushes the sharpening front over to the cyclonic side, leaving it in a region of surface convergence that accelerates its sharpening. The front sharpens super-exponentially and collapses into a discontinuity in finite time. While these austere models of frontogenesis do not capture the fact that fronts are embedded in a turbulent meso- and submesoscale eddy field, and they neglect the impact of mixed-layer turbulence, the dynamics are transparent and well-understood, offering clear insight into how kinetic energy is transferred downscale at fronts.

We quantify the scale-resolved kinetic energy budget of frontogenesis using a spectral budget and a coarse-graining approach. The spectral budget quantifies the net role of downscale kinetic energy transfer in energizing the sharpening along-front jet, relative to buoyancy production and other terms in the budget (Section~\ref{sec:spectral}). The coarse-graining approach further localizes scale transfers, revealing geostrophic downscale transfer on both flanks of the front and a dipole of ageostrophic transfer, with a downscale transfer on the cyclonic side of the front and an upscale transfer on the anticyclonic side (Section~\ref{sec:coarse}). We discuss how these results compare to more realistic flows and highlight the limitations of the simple models studied here (Section~\ref{sec:discussion}).

\section{Frontogenesis: evolution and energy budget}
\label{sec:equations}

This section introduces the frontal setup and reviews the frontal evolution under both QG and SG dynamics, closely following \textcite{stone1966,hoskins72}.
We derive kinetic energy equations in physical and spectral space, and we apply the coarse-graining framework to the frontal setup.
Readers already familiar with the QG and SG dynamics of a strain-driven front may skip Sections \ref{sec:qg} and~\ref{sec:sg} and proceed to Sections \ref{sec:spec_eqn} and~\ref{sec:coarse_eqn} for the discussion of scale-resolved energy budgets, although we encourage them to review the kinetic energy budgets \eqref{eq:qg_ke} and~\eqref{eq:sg_ke} before proceeding.

\subsection{Frontal setup}
\label{sec:frontal_setup}

We consider classic strain-driven frontogenesis. A strain field $U = -\alpha x$ and $V = \alpha y$, with a fixed strain rate~$\alpha$, is imposed and acts on a pre-existing buoyancy gradient in the $x$-direction. We consider a front that is initially broad and sharpens exponentially at early times, before the frontal circulation feeds back on its evolution. We prescribe this initial condition as
\begin{equation}
  b(x, z, t) \to N^2 z + B(x e^{\alpha t}) = N^2 z + \frac{\Delta}{2} \tanh \frac{x e^{\alpha t}}{\ell} \quad \text{as} \quad t \to -\infty,
\end{equation}
where $\Delta$ is the buoyancy difference across the front. This initial condition is chosen such that the buoyancy field would be $b = N^2 z + B(x)$ at $t = 0$ if the strain field was all that acted on it. We set the nominal frontal width at that time to $\ell = NH/f$ without loss of generality because a change in the frontal width is equivalent to a shift of the time coordinate.
The initial flow is chosen to be in balance, such that no unbalanced flow is excited: $u \to -2 \alpha z e^{\alpha t} B'(x e^{\alpha t})/f^2$, $v \to z e^{\alpha t} B'(x e^{\alpha t})/f$ as $t \to -\infty$.
This initial condition also implies that the potential vorticity (PV) is uniform and remains so for all times.
Furthermore, the buoyancy field is independent of~$y$ and remains so over the course of the evolution because the frontal flow, the departure from the imposed strain field, is also independent of~$y$. Frontal instabilities are not allowed to develop.

\begin{figure}[t]
  \centering
  \includegraphics[width=\textwidth]{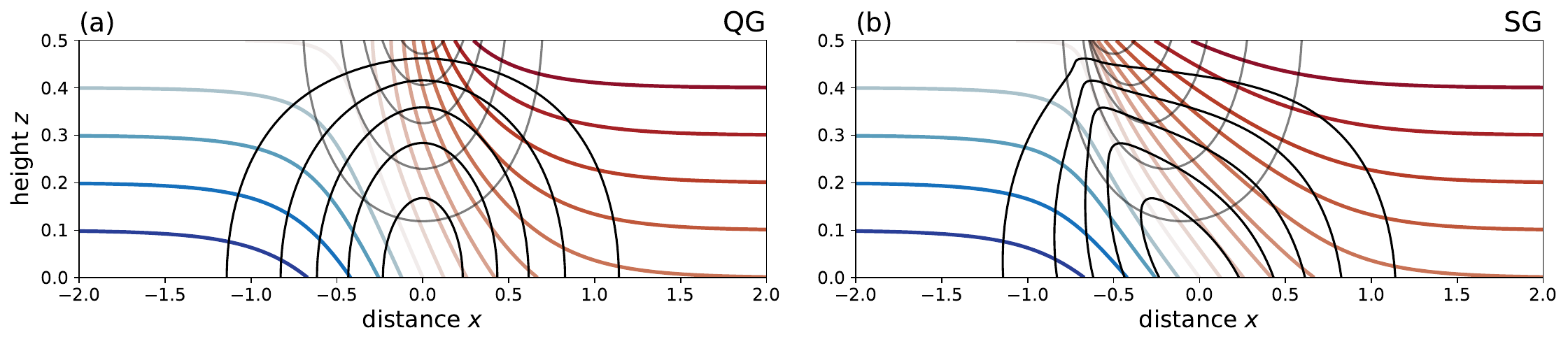}
  \caption{Quasi-geostrophic and semi-geostrophic fronts. (a)~The buoyancy field $N^2z + b_0$ (color), the along-front flow~$v_0$ (gray), and the ageostrophic cross-frontal circulation~$\chi$ (black) for the QG solution. (b)~The buoyancy field~$b$ (color), the along-front flow~$v_g$ (gray), and the ageostrophic cross-frontal circulation~$\chi$ (black) for the SG solution. The fields are shown at $t = \num{1.33}$ and for $\gamma = 1$. Note that only the top half of the domain is shown. All variables are dimensionless (see Appendix~\ref{sec:nondim}).}
  \label{fig:config}
\end{figure}

We proceed by listing the momentum and buoyancy equations that describe the evolution of the front under QG and SG assumptions. In both cases, the dynamics are best cast in terms of buoyancy advection at the boundaries and an inversion statement that determines the flow given the field of buoyancy at the boundaries. We also state the respective omega equations that allow us to explicitly calculate the ageostrophic cross-frontal circulation, and we derive the respective kinetic energy equations. While the cross-frontal circulation is not needed to determine the evolution, it plays an important role in the dynamics. Horizontal advection by the ageostrophic circulation, neglected in QG and retained in SG, distinguishes the two theories. While the front remains symmetric in QG, it is tilted over by the ageostrophic circulation in SG, and frontogenesis is accelerated by the horizontal convergence of the ageostrophic flow on the cyclonic side of the front (Fig.~\ref{fig:config}), leading to frontal collapse in finite time.

\subsection{Frontal evolution and kinetic energy in QG}
\label{sec:qg}

Following the usual approach to QG dynamics \parencite[e.g.,][]{pedlosky1987}, we expand the flow variables of the Boussinesq equations in a perturbation series in a Rossby number $\varepsilon \ll 1$: the cross-front velocity $u = u_0 + \varepsilon u_1 + \ldots$, the along-front velocity $v = v_0 + \varepsilon v_1 + \ldots$, the vertical velocity $w = w_0 + \varepsilon w_1 + \ldots$, the buoyancy $b = N^2 z + b_0 + \varepsilon b_1 + \ldots$, and the geopotential $\phi = \frac{1}{2} N^2 z^2 + \phi_0 + \varepsilon \phi_1 + \ldots$ The buoyancy and geopotential~$\phi = p/\rho_0$ include background terms that represent a hydrostatically balanced constant stratification~$N^2$. All variables describe departures from the underlying strain field, which is in balance with a geopotential field $-\frac{1}{2} \alpha^2 (x^2 + y^2) + f \alpha x y$. In QG scaling, the background stratification is assumed to be strong, such that the zeroth-order buoyancy equation implies $w_0 = 0$. The zeroth-order momentum equations are
\begin{align}
  -fv_0 &= -\frac{\partial \phi_0}{\partial x}, \label{eq:qg-0-x} \\
  fu_0 &= 0, \label{eq:qg-0-y} \\
  0 &= -\frac{\partial \phi_0}{\partial z} + b_0; \label{eq:qg-0-z}
\end{align}
and the first-order equations are
\begin{align}
  -f v_1 &= -\pp{\phi_1}{x}, \label{eq:qg-1-x} \\
  \pp{v_0}{t} - \alpha x \pp{v_0}{x} + \alpha v_0 + f u_1 &= 0, \label{eq:qg-1-y} \\
  0 &= -\frac{\partial \phi_1}{\partial z}+b_1. \label{eq:qg-1-z}
\end{align}
The advection terms in~\eqref{eq:qg-1-y} arise from the strain field advecting the geostrophic along-front geostrophic flow~$v_0$ and the along-front geostrophic flow~$v_0$ advecting the strain field. 

The assumption of a uniform interior PV implies that the QGPV vanishes identically, such that the evolution is fully governed by the advection of zeroth-order buoyancy at the top and bottom boundaries:
\begin{equation}
  \frac{\partial b_0}{\partial t} - \alpha x \frac{\partial b_0}{\partial x}  = 0 \quad \text{at} \quad z = \pm \frac{H}{2}.
  \label{eq:qg_buoyancy}
\end{equation}
Given the absence of along-front buoyancy gradients and of a cross-front geostrophic flow, buoyancy is advected by the strain field only, and the initial buoyancy gradient is sharpened exponentially:
\begin{equation}
  b_0(x, z, t) = B(x e^{\alpha t}) \quad \text{at} \quad z = \pm \frac{H}{2}.
\end{equation}
This buoyancy solution at the boundaries can be applied at any time~$t$ and provides the boundary conditions for the inversion relation that can be solved for the zeroth-order geopotential~$\phi_0$:
\begin{equation}
  \frac{\partial^2 \phi_0}{\partial x^2} + \frac{f^2}{N^2} \frac{\partial^2 \phi_0}{\partial z^2} = 0, \quad \pp{\phi_0}{z} = B(x^{\alpha t}) \quad \text{at} \quad z = \pm \frac{H}{2}.
  \label{eq:qg_balance}
\end{equation}
The streamfunction then determines the interior buoyancy field~$b_0 = \partial_z \phi_0$ and the along-front geostrophic flow~$v_0 = f^{-1} \partial_x \phi_0$.

The sharpening of the buoyancy gradient by the strain field induces an ageostrophic cross-frontal circulation. In the $x$--$z$~plane, we define the streamfunction $\chi$ of the cross-frontal circulation,
\begin{equation}
  u_1 = -\frac{\partial \chi}{\partial z} \quad \text{and} \quad w_1 = \frac{\partial \chi}{\partial x},
\end{equation}
and diagnose it using the omega equation
\begin{equation}
  N^2 \frac{\partial^2 \chi}{\partial x^2} + f^2 \frac{\partial^2 \chi}{\partial z^2} = 2\alpha \frac{\partial b_0}{\partial x}.
  \label{eq:qg_omega}
\end{equation}
As usual, the vertical advection by this ageostrophic circulation is taken into account implicitly in the evolution of the front as described by buoyancy advection on the boundaries and the inversion relation. We solve all equations in Fourier space in the horizontal, using discrete sine and cosine transforms, and we apply an analytical vertical structure for each horizontal wavenumber.

The rigid-lid condition $w_1 = 0$ at $z = \pm H/2$ allows the strain field to sharpen the front most rapidly there (Fig.~\ref{fig:config}a, colored contours). An along-front geostrophic flow~$v_0$ develops with its maximum at the surface and center of the domain (Fig.~\ref{fig:config}a, gray). In the interior, the sharpening of the cross-frontal buoyancy gradient induces an ageostrophic circulation (Fig.~\ref{fig:config}a, black) that consists of downwelling on the dense side of the front and upwelling on the light side. This acts to flatten the cross-frontal buoyancy gradient and counteracts its sharpening by the strain field in the interior, such that the interior buoyancy gradient remains modest. This QG frontal evolution remains symmetric around $x = 0$ because only the strain field advects buoyancy and the geostrophic along-front momentum.

The kinetic energy budget involves both zeroth-order and first-order quantities, diagnosed from the inversion relation~\eqref{eq:qg_balance} and omega equation~\eqref{eq:qg_omega}. The equation for the geostrophic kinetic energy $\frac{1}{2} v_0^2$ is formed by taking the sum of \eqref{eq:qg-0-x}--\eqref{eq:qg-0-z} dotted with the first-order velocity vector~$(u_1, v_1, w_1)$ and \eqref{eq:qg-1-x}--\eqref{eq:qg-1-z} dotted with the zeroth-order velocity vector $(0, v_0, 0)$:
\begin{equation}
  \pp{}{t} \frac{v_0^2}{2} - \alpha x \pp{}{x} \frac{v_0^2}{2} + \alpha v_0^2 + u_1 \pp{\phi_0}{x} + w_1 \pp{\phi_0}{z} = w_1 b_0. 
\label{eq:qg_mom}
\end{equation}
Written in flux form, taking advantage of the non-divergence of $(u_1, w_1)$ in the $x$--$z$~plane, this becomes
\begin{equation}
  \pp{}{t} \frac{v_0^2}{2} + \pp{}{x} \bigg( -\alpha x \frac{v_0^2}{2} + u_1 \phi_0 \bigg) + \pp{}{z} \bigg( w_1 \phi_0 \bigg) + \frac{3}{2} \alpha v_0^2 = w_1 b_0.
  \label{eq:qg_ke}
\end{equation}
This equation indicates that there is a sink~$-\frac{3}{2} \alpha v_0^2$ of geostrophic kinetic energy density that arises from a combination of the strain field stretching out the front in the $y$-direction and the shear production term~$-\alpha v_0^2$ \parencite[cf.,][]{shakespeare_generation_2015}. It is important to treat this term as a sink rather an advective redistribution. The redistribution terms, in contrast, vanish upon integration over $x$ and~$z$ as long as the front is sufficiently confined, as is the case in the example considered in this paper. 

\subsection{Frontal evolution and kinetic energy in SG}
\label{sec:sg}

Following \textcite{hoskins72}, we now take into account the horizontal advection by the ageostrophic cross-frontal circulation by making the geostrophic momentum approximation rather than the QG approximation. We define the geostrophic velocities as
\begin{equation}
  v_g \equiv \frac{1}{f}\frac{\partial \phi}{\partial x} \quad \text{and} \quad u_g \equiv -\frac{1}{f}\frac{\partial \phi}{\partial y} = 0,
\end{equation}
and we write the SG momentum equations as
\begin{align}
  -f v &= -\pp{\phi}{x}, \label{eq:sg_mom_x} \\
  \frac{\partial v_g}{\partial t} + (-\alpha x+u)\frac{\partial v_g}{\partial x} + w\frac{\partial v_g}{\partial z} + \alpha v_g + fu &= 0, \label{eq:sg_mom_y} \\
  0 &= -\pp{\phi}{z} + b \label{eq:sg_mom_z}.
\end{align}
We note that \eqref{eq:sg_mom_x} implies that $v = v_g$. The buoyancy equation also now includes the full ageostrophic advection:
\begin{equation}
  \frac{\partial b}{\partial t} + (-\alpha x +u)\frac{\partial b}{\partial x} + w\frac{\partial b}{\partial z} = 0.
  \label{eq:sg_mom_b}
\end{equation}
These equations for the evolution of a strain-driven front are most easily solved by performing a coordinate transformation to geostrophic coordinates:
\begin{equation}
  X = x + \frac{v_g}{f}, \quad Z = z, \quad T = t.
  \label{eq:sg_geo_coord}
\end{equation}
We further define the modified geopotential
\begin{equation}
  \Phi \equiv \phi + \frac{v_g^2}{2},
\end{equation}
such that
\begin{equation}
  f v_g = \frac{\partial \phi}{\partial x} = \frac{\partial \Phi}{\partial X}, \quad b = \frac{\partial \phi}{\partial z} = \frac{\partial \Phi}{\partial Z},
\end{equation}
in close analogy with the leading-order balances in QG. Based on this coordinate transformation, the along-front momentum and buoyancy equations \eqref{eq:sg_mom_y} and~\eqref{eq:sg_mom_b} simplify to: 
\begin{align}
  \pp{v_g}{T} - \alpha X \pp{v_g}{X} + w \pp{v_g}{Z} + \alpha v_g + fu &= 0, \label{eq:sg-geo-y} \\
  \pp{b}{T} - \alpha X \pp{b}{X} + w \pp{b}{Z} &= 0. \label{eq:sg-geo-b}
\end{align}
Notably, horizontal advection is only by the strain field in geostrophic coordinates, just like in QG in physical space. At $z = \pm H/2$, where $w = 0$, the buoyancy equation takes the same form as in QG. The perturbation buoyancy~$b' \equiv b - N^2 z$ can thus be solved for at the boundaries as before:
\begin{equation}
  b'(X, Z, T) = B(Xe^{\alpha T}) \quad \text{at} \quad Z = \pm \frac{H}{2}.
\end{equation}
Furthermore, the inversion relation takes the same form as in QG as well, such that the evolution is entirely analogous except that it occurs in geostrophic coordinates---the ageostrophic advection is implicit in the coordinate transformation back to physical space.

The SG inversion relation in geostrophic coordinates is
\begin{equation}
  \frac{1}{f^2}\frac{\partial^2 \Phi}{\partial X^2} + \frac{f}{q_g}\frac{\partial^2 \Phi}{\partial Z^2} = 1, 
\end{equation}
in which
\begin{equation}
  q_g = \left( f + \pp{v_g}{x} \right) \pp{b}{z} - \pp{v_g}{z} \pp{b}{x} = f N^2
\end{equation}
is the geostrophic potential vorticity, which is equal to the uniform value of the unperturbed fluid. Conservation of geostrophic potential vorticity implies that $q_g$ will remain at this value for all time. If we further split $\Phi$ into a background and a perturbation component,
\begin{equation}
    \Phi = \frac{1}{2}N^2 z^2 + \Phi',
\end{equation}
the inversion relation for the perturbation becomes
\begin{equation}
  \frac{\partial^2 \Phi^\prime}{\partial X^2} + \frac{f^2}{N^2}\frac{\partial^2 \Phi^\prime}{\partial Z^2} = 0, \quad \pp{\Phi'}{z} = B(X e^{\alpha T}) \quad \text{at} \quad Z = \pm \frac{H}{2},
\end{equation}
which has the same form as the QG inversion relation in~\eqref{eq:qg_balance}.

The SG omega equation is derived following \textcite{hoskins1977}. We write \eqref{eq:sg-geo-y} and~\eqref{eq:sg-geo-b} as 
\begin{align}
  \pp{v_g}{T} - \alpha X \pp{v_g}{X} + \alpha v_g + f \tilde u &= 0, \label{eq:sg-geo-star-y} \\
  \pp{b}{T} - \alpha X \pp{b}{X} + \tilde w N^2 &= 0, \label{eq:sg-geo-star-b}
\end{align}
where
\begin{equation}
  \tilde u = u + \frac{w}{f}\frac{\partial v_g}{\partial Z}, \quad \tilde w = \frac{w}{J}, \quad \text{and} \quad J \equiv 1 + \frac{1}{f} \pp{v_g}{x} = \bigg( 1 - \frac{1}{f} \pp{v_g}{X} \bigg)^{-1}.
  \label{eq:sg-star}
\end{equation}
These redefined velocities in the cross-front plane are also nondivergent, so we obtain the SG omega equation
\begin{equation}
  N^2 \pp{^2 \tilde \chi}{X^2} + f^2 \pp{^2 \tilde \chi}{Z^2} = 2\alpha \pp{b}{X}
\end{equation}
for the cross-front streamfunction of the redefined velocities,
\begin{equation}
  \tilde u = -\pp{\tilde \chi}{Z}, \quad \tilde w = \pp{\tilde \chi}{X}.
\end{equation}
Aside from being formulated in geostrophic coordinates and for the redefined velocities, this has again the same form as the QG omega equation, so QG solutions to the frontal evolution can be translated to SG by applying the transformation~\eqref{eq:sg-star} and the coordinate transformation~\eqref{eq:sg_geo_coord}.

In physical space, the frontal evolution in SG (Fig.~\ref{fig:config}b) differs distinctly from the QG case (Fig.~\ref{fig:config}a). The coordinate transformation in~\eqref{eq:sg_geo_coord}, which encodes the ageostrophic advection, slumps the surface front toward the dense cyclonic side. The core of the front, where the along-front geostrophic velocity~$v_g$ is the strongest (Fig.~\ref{fig:config}b, grey), is also shifted. The cyclonic side of the front is compressed, while the anticyclonic side is stretched. 
This slumping further places the front in a region in which the ageostrophic cross-frontal circulation is horizontally convergent, which accelerates the sharpening of the surface buoyancy gradient (Fig.~\ref{fig:config}b, color) and causes frontal collapse in finite time.
In the interior, the ageostrophic circulation (Fig.~\ref{fig:config}b, black) is also visibly distorted compared with the QG solution. It is characterized by a stronger and more concentrated downwelling on the dense cyclonic side of the front, while the upwelling becomes weaker and more expansive on the light anticyclonic side. 

With the solution for the geostrophic and ageostrophic flow as well as the buoyancy field in hand, we can diagnose all terms in the SG kinetic energy budget. We form this budget by dotting \eqref{eq:sg_mom_x}--\eqref{eq:sg_mom_z} with the flow $(u, v, w)$:
\begin{equation}
  \pp{}{t} \frac{v_g^2}{2} + \pp{}{x} \bigg[ (-\alpha x + u) \frac{v_g^2}{2} + u \phi \bigg] + \pp{}{z} \bigg[ w \frac{v_g^2}{2} + w \phi \bigg] + \frac{3}{2} \alpha v_g^2 = w b.
  \label{eq:sg_ke}
\end{equation}
This equation, in comparison with the QG budget~\eqref{eq:qg_mom}, includes the additional advection terms of geostrophic kinetic energy by the ageostrophic velocities $(u, w)$. This ageostrophic advection contributes to the scale transfer of kinetic energy, as discussed next.

\subsection{Spectral kinetic energy budgets}
\label{sec:spec_eqn}

The spectral energy budgets are derived following a standard procedure. First, a Fourier transform is applied to the momentum equations. The transformed equations are then dotted with the complex conjugate of the Fourier transform of the velocity vector, analogous to the formation of the kinetic energy equations in physical space discussed above. Taking the real part then yields the respective kinetic energy equation. For QG,
\newcommand{\vp}{\vphantom{\pp{}{z}}}
\begin{equation}
  \underbracket[0.16ex]{\pp{}{t} \frac{|\hat{v}_0|^2}{2} \vp}_{\text{tendency}} 
  \underbracket[0.16ex]{- \, \Re \hat{v}_0^* \widehat{\alpha x \pp{v_0}{x}} - \frac{\alpha}{2}|\hat{v}_0|^2 \vp}_{\text{geostrophic advection}}
  \underbracket[0.16ex]{+ \, k \Im \hat{u}_1^*\hat{\phi}_0 + \Re \pp{}{z} \left( \hat{w}_1^*\hat{\phi}_0 \right)}_{\text{ pressure work}}
  \underbracket[0.16ex]{+ \, \frac{3}{2} \alpha |\hat{v}_0|^2 \vp}_{\text{sink}}
  =
  \underbracket[0.16ex]{\Re \hat{w}_1^*\hat{b}_0 \vp}_{\text{buoy.\ prod.}},
\label{eq:qg_spec}
\end{equation}
and for SG,
\renewcommand{\vp}{\vphantom{\pp{}{z}}}
\begin{multline}
  \underbracket[0.16ex]{\pp{}{t} \frac{|\hat{v}_g|^2}{2} \vp}_{\text{tendency}} 
  \underbracket[0.16ex]{- \, \Re \hat{v}_g^* \widehat{\alpha x \frac{\partial v_g}{\partial x}} - \frac{\alpha}{2}|\hat{v}_g|^2 \vp}_{\text{geostrophic advection}} 
   \underbracket[0.16ex]{+ \, \Re \hat{v}_g^* \widehat{u \pp{v_g}{x}} + \Re \hat{v}_g^* \widehat{w \pp{v_g}{z}} \vp}_{\text{agostrophic advection}} \\
  \underbracket[0.16ex]{+ \, k\Im \hat{u}^* \hat{\phi} + \Re\pp{}{z} \hat{w}^*\hat{\phi}}_{\text{ pressure work}} 
  \underbracket[0.16ex]{+ \, \frac{3}{2} \alpha |\hat{v}_g|^2 \vp}_{\text{sink}}
  =
  \underbracket[0.16ex]{\Re \hat{w}^*\hat{b} \vp}_{\text{buoy.\ prod.}}.
\label{eq:sg_spec}
\end{multline}
The terms are annotated with the corresponding physical process. As in the budgets in physical space, the sink terms arise from a combination of the along-front stretching and shear production involving the strain field. These processes dilute the kinetic energy density of the front and should be treated as sinks rather than advective scale transfers.

We write the SG spectral energy budget in physical rather than geostrophic coordinates to understand the role of ageostrophic advection and analyze the transfer across the actual spatial scales of the front. In geostrophic coordinates, where the horizontal advection by the ageostrophic flow is implicit, the SG energy budget would take a form very similar to the QG budget, with the exception that vertical advection of geostrophic momentum would still remain. The budget in physical coordinates also allows a more direct comparison with previous diagnostics from realistic primitive-equation simulations as well as observations, which invariably operate in physical coordinates, where ageostrophic advection has been singled out to play an outsized role.

The advection terms in \eqref{eq:qg_spec} and~\eqref{eq:sg_spec} redistribute energy and can be written as the divergence of a spectral flux~$\Pi(k)$, defined as
\begin{equation}
  \Pi(k) = \int_0^k A(\kappa) \, \mathrm{d} \kappa. 
  \label{eq:spec_flux}
\end{equation}
Here, $A(k)$ denotes the advection term under consideration, geostrophic or ageostrophic, QG or SG, as written on the left-hand sides of \eqref{eq:qg_spec} and~\eqref{eq:sg_spec}. If $\Pi(k) > 0$, energy is transferred from scales larger than $k^{-1}$ to scales smaller than $k^{-1}$, which we refer to as a downscale transfer. Conversely, a negative $\Pi(k)$ implies an upscale transfer from small to large scales.\footnote{That the ageostrophic terms in SG can only redistribute energy across horizontal scales at each level follows from the fact that $\int w \frac{v_g^2}{2} \, \d x = \int \tilde w \frac{v_g^2}{2} J \, \d x = \int \tilde w \frac{v_g^2}{2} \, \d X = 0$ by the symmetry of the solution in geostrophic coordinates. There is no advective transport of kinetic energy in the vertical.}

\subsection{Coarse-grained kinetic energy budget}
\label{sec:coarse_eqn}

Coarse-grained kinetic energy budgets are derived to assess the spatial distribution of kinetic energy transfer in physical space. The derivation closely follows \textcite{aluie2018}, where additional detail can be found. Here, we specialize the approach to the frontal evolution with a prescribed strain field. We decompose fields into scales larger and smaller than a spatial filter length~$L$ using a Gaussian filter. Filtered fields are denoted with an overbar. We begin with a general discussion of the coarse-graining technique applied to departures from a prescribed mean flow. The resulting coarse-grained kinetic energy balance is then applied specifically to the QG and SG frontal cases.

The Boussinesq momentum equation governing the evolution of the flow deviation from a prescribed mean flow is:
\begin{equation}
  \pp{u_i}{t} + (U_j + u_j) \pp{u_i}{x_j} + u_j \pp{U_i}{x_j} + f e_{ijk} z_j u_k = -\pp{\phi}{x_i} + b z_i,
\label{eq:coarse_mom}
\end{equation}
with $U_i$ representing the mean flow, $u_i$ the departure from the mean, $z$ the vertical unit vector, and $e_{ijk}$ the Levi-Civita symbol. Summation over repeated indices is implied. Following \textcite{aluie2018}, we apply the filter to~\eqref{eq:coarse_mom} before multiplying it by the filtered velocity $\mean{u}_i$. After some rearranging, one obtains the following coarse-grained kinetic energy budget:
\renewcommand{\vp}{\vphantom{\bigg[ \pp{}{x_j} (\mean{U}_j + \mean{u}_j) \frac{\mean{u}_i^2}{2} \bigg]}}
\begin{multline}
  \underbracket[0.16ex]{\pp{}{t} \frac{\mean{u}_i^2}{2} \vp}_\text{tendency}
  \underbracket[0.16ex]{+ \, \pp{}{x_j} \bigg[ (\mean{U}_j + \mean{u}_j) \frac{\mean{u}_i^2}{2} \bigg] \vp}_\text{resolved advection}
  \underbracket[0.16ex]{+ \, \pp{}{x_j} \bigg[ \mean{u}_i \bigg( \mean{U_j u_i} - \mean{U}_j \mean{u}_i + \mean{u_j U_i} - \mean{u}_j \mean{U}_i + \mean{u_j u_i} - \mean{u}_j \mean{u}_i \bigg) \bigg] \vp}_\text{subgrid advection} \\
  \underbracket[0.16ex]{- \, \pp{\mean{u}_i}{x_j} \bigg[ \mean{U_j u_i} - \mean{U}_j \mean{u}_i + \mean{u_j U_i} - \mean{u}_j \mean{U}_i + \mean{u_j u_i} - \mean{u}_j \mean{u}_i \bigg] \vp}_\text{scale transfer $\Pi$}
  \underbracket[0.16ex]{+ \, \mean{u}_i \mean{u}_j \pp{\mean{U}_i}{x_j}}_\text{shear prod.}
  \underbracket[0.16ex]{+ \, \pp{}{x_i} \Big( \mean{u}_i \mean{\phi} \Big) \vp}_\text{pressure work}
  = \underbracket[0.16ex]{\mean{w} \mean{b}, \vp}_\text{buoy.\ prod.}
\label{eq:coarse_general}
\end{multline}
where~$w = u_i z_i$ is the vertical velocity. Each term is annotated with the corresponding physical process. The sum of the terms labeled ``resolved advection,'' ``subgrid advection,'' ``shear production,'' and ``pressure work'' corresponds to the divergence of the transport terms defined in \textcite{aluie2018}. We note that $\Pi$ in this context depends on both the filter scale~$L$ and the position~$x$; it must be integrated over~$x$ to obtain the scale transfer analogous to the $\Pi(k)$ in the spectral budget.

To apply~\eqref{eq:coarse_general} to the QG front, we identify the background strain field components $-\alpha x$ and $\alpha y$ with the mean flow $U_i$, while the geostrophic velocity~$v_0$ and the ageostrophic velocity components~$(u_1, w_1)$ correspond to the fluctuation field $u_i$. Because the ageostrophic velocities do not participate in momentum advection in QG dynamics, only the geostrophic component contributes to the advective terms. We then obtain the following form of the coarse-grained QG kinetic energy equation for the developing front:
\renewcommand{\vp}{\vphantom{\bigg[\frac{\mean{v}_0^2}{2} \bigg]}}
\begin{multline}
  \underbracket[0.16ex]{\pp{}{t} \frac{\mean{v}_0^2}{2} \vp}_{\text{tendency}} 
  \underbracket[0.16ex]{+ \, \pp{}{x} \bigg[ (\overline{-\alpha x}) \frac{\mean{v}_0^2}{2} \bigg]}_{\text{resolved advection}}
  \underbracket[0.16ex]{+ \, \pp{}{x} \bigg( \mean{v}_0 \big[ \overline{v_0(-\alpha x)} - \mean{v}_0(\overline{-\alpha x}) \big] \bigg) \vp}_{\text{subgrid advection}} \\
  \underbracket[0.16ex]{- \, \frac{\partial \mean{v}_0}{\partial x} \bigg[ \overline{v_0(-\alpha x)} - \mean{v}_0 (\overline{-\alpha x}) \bigg] \vp}_{\text{scale transfer $\Pi$}}
  \underbracket[0.16ex]{+ \, \frac{3}{2}\alpha \mean{v}_0^2 \vp}_{\text{sink}}
  \underbracket[0.16ex]{+ \, \pp{}{z}(\mean{w}_1\mean{\phi}_0) + \pp{}{x}(\mean{u}_1\mean{\phi}_0) \vp}_{\text{pressure work}} 
= \underbracket[0.16ex]{\mean{w}_1\mean{b}_0\vp}_{\text{buoy.\ prod.}}.
\label{eq:qg_coarse}
\end{multline}
We note that $\mean{x} = x$.

Similarly, the coarse-grained equation for SG dynamics is derived based on \eqref{eq:coarse_general}, with the distinction that the SG geostrophic momentum is additionally advected by the ageostrophic velocities. The coarse-grained SG kinetic energy equation for the strain-driven front is therefore:
\renewcommand{\vp}{\vphantom{\pp{}{x} \bigg[ \bigg]}}
\begin{multline}
  \underbracket[0.16ex]{\pp{}{t} \frac{\mean{v}_g^2}{2} \vp}_{\text{tendency}}
  \underbracket[0.16ex]{+ \, \pp{}{x} \bigg[ \overline{(-\alpha x + u)} \frac{\mean{v}_g^2}{2} \bigg] + \pp{}{z} \bigg( \mean{w}\frac{\mean{v}_g^2}{2} \bigg)}_{\text{resolved advection}} \\
  \underbracket[0.16ex]{+ \, \pp{}{x} \bigg( \mean{v}_g \big[ \overline{v_g (-\alpha x + u)} - \mean{v}_g \overline{(-\alpha x + u)} \big] \bigg) + \pp{}{z} \bigg[ \mean{v}_g(\overline{v_gw}-\mean{v}_g\mean{w}) \bigg]}_{\text{subgrid advection}} \\
  \underbracket[0.16ex]{- \, \pp{\mean{v}_g}{x} \bigg[ \overline{v_g (-\alpha x)}-\mean{v}_g\overline{(-\alpha x)} \bigg] - \pp{\mean{v}_g}{x} (\overline{v_g u} - \mean{v}_g\mean{u}) - \pp{\mean{v}_g}{z} (\overline{v_g w} - \mean{v}_g\mean{w})}_{\text{scale transfer $\Pi$}} \\
  \underbracket[0.16ex]{+ \, \frac{3}{2}\alpha \mean{v}_g^2 \vp}_{\text{sink}} 
  \underbracket[0.16ex]{+ \, \pp{}{z}(\mean{w}\mean{\phi}) + \pp{}{x}(\mean{u}\mean{\phi}) \vp}_{\text{pressure work}} 
= \underbracket[0.16ex]{\mean{w}\mean{b}.\vp}_{\text{buoy.\ prod.}}
\label{eq:sg_coarse}
\end{multline}
The resolved kinetic energy is not only transported by the resolved flow, but the subgrid flow also effects such a transport (labeled ``subgrid advection''). Note that the sink terms $\frac{3}{2}\alpha \mean{v}_0^2$ in QG and $\frac{3}{2}\alpha \mean{v}_g^2$ in SG again stem from shear production and the resolved advection in the along-front direction in~\eqref{eq:coarse_general}, the same as in physical and spectral space. We further distinguish between the geostrophic (first term) and ageostrophic (second and third terms) scale transfers in SG.

\subsection{Nondimensionalization}
\label{sec:nondim_short}

All analysis will be presented in dimensionless form. We use the following scales to nondimensionalize the QG equations:
\begin{equation}
  t \sim \alpha^{-1}, \quad x \sim \frac{N H}{f}, \quad z \sim H, \quad v_0 \sim \frac{\Delta}{N}, \quad u_1 \sim \frac{\alpha \Delta}{f N} \quad w_1 \sim \frac{\alpha \Delta}{N^2}, \quad \phi_0 \sim \Delta H, \quad b_0 \sim \Delta.
  \label{eqn:nondimqg}
\end{equation}
This eliminates all dimensional factors and leaves the frontal evolution independent of any dimensionless parameters. In the SG case, we apply the equivalent scaling:
\begin{equation}
  t \sim \alpha^{-1}, \quad x \sim \frac{N H}{f}, \quad z \sim H, \quad v_g \sim \frac{\Delta}{N}, \quad u \sim \frac{\alpha \Delta}{f N} \quad w \sim \frac{\alpha \Delta}{N^2}, \quad \phi \sim \Delta H, \quad b \sim \Delta.
  \label{eqn:nondimsg}
\end{equation}
The dimensionless SG equations depend on the parameter~$\gamma \equiv \Delta/N^2 H$, which multiplies the ageostrophic advection terms and also appears in the Jacobian of the coordinate transformation to geostrophic coordinates (see Appendix~\ref{sec:nondim} for details). This parameter is the ratio between the buoyancy drop~$\Delta$ across the front and the vertical buoyancy contrast~$N^2 H$. Equivalently, it measures the vertical excursion~$\Delta/N^2$ of isopycnals across the front relative to the domain height~$H$ (Fig.~\ref{fig:config}). The QG scaling assumes $\gamma \ll 1$. In the following, we consider the SG case $\gamma = 1$ in detail and examine the $\gamma$~dependence in Section~\ref{sec:gamma}. From hereon, all quantities are understood to be nondimensionalized, and we omit decorations for simplicity.

\section{Energy budget analysis}
\label{sec:energy}

We now apply the energy diagnostics developed in the previous section to the frontal evolution at a time close to the collapse of the SG front ($t = \num{1.33}$ for $\gamma = 1$, see Fig.~\ref{fig:gamma}a). We assess the importance of the various physical processes in the spectral energy budget and highlight the role of ageostrophic advection in transferring energy to small scales in the SG case. We then apply the coarse-graining approach to further investigate the spatial pattern of cross-scale energy transfer and exhibit the asymmetry of these transfers between the cyclonic and anticyclonic sides of the front.

\subsection{Spectral kinetic energy budgets}
\label{sec:spectral}

\begin{figure}[t]
  \centering
  \includegraphics[width=\textwidth]{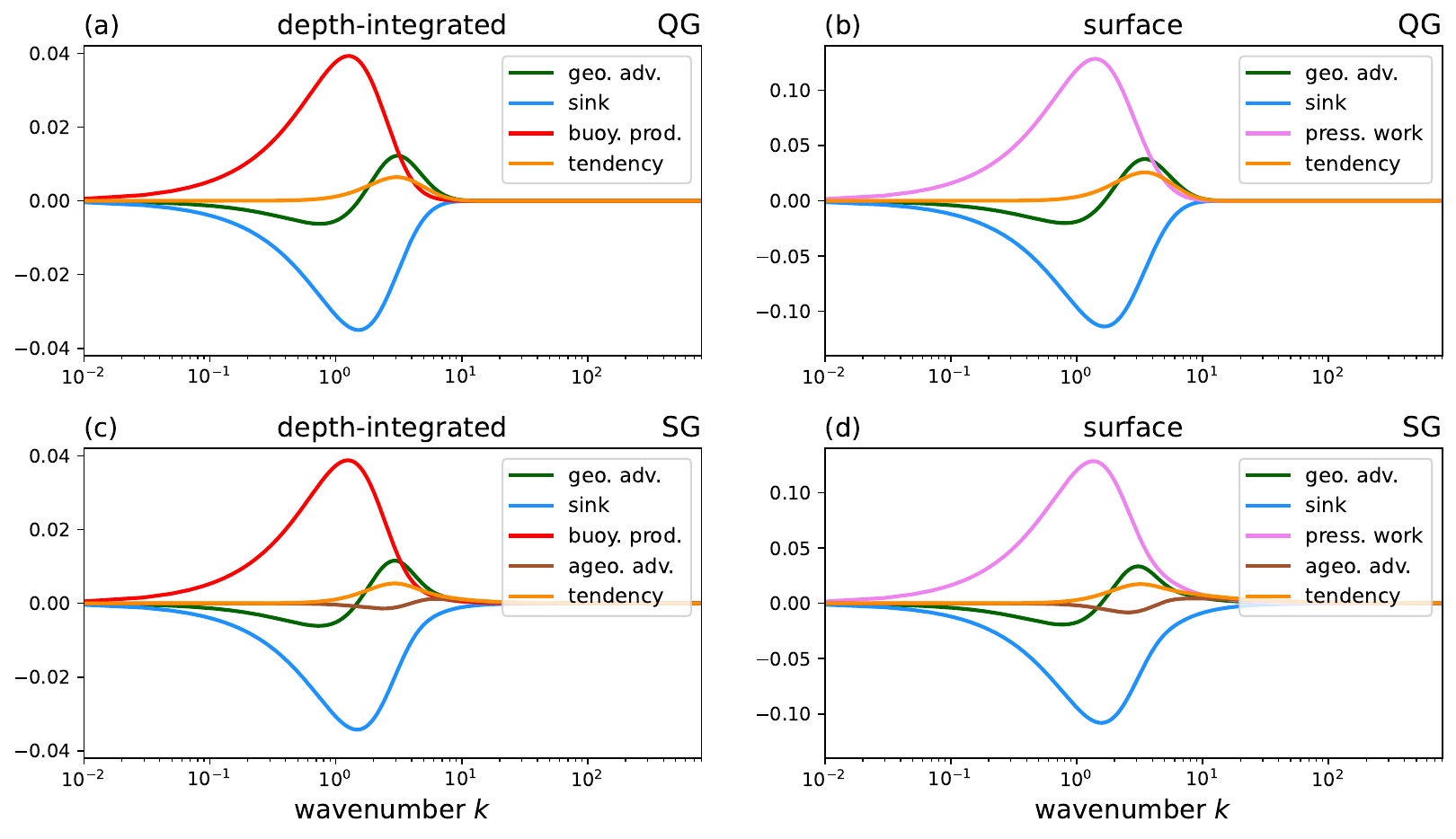}
  \caption{Spectral kinetic energy budget for the QG and SG fronts ($\gamma = 1$). The terms in~\eqref{eq:qg_spec} and~\eqref{eq:sg_spec} are shown with a sign convention as if all terms except the tendency were on the right-hand sides of these equations. Under this convention, negative values indicate energy removal and positive values indicate energy gain. Both geostrophic and ageostrophic advection transfer energy from large to small spatial scales. The budgets are evaluated at $t = \num{1.33}$, shortly before the collapse of the SG front. (a,c)~Budgets integrated over the depth of the domain. (b,d)~Budgets at the surface ($z = \frac{1}{2}$). All variables are dimensionless (see Appendix~\ref{sec:nondim}).}
  \label{fig:spec}
\end{figure}

The spectral kinetic energy budgets for QG and SG frontogenesis exhibit strong similarities (Fig.~\ref{fig:spec}). Both cases show a transfer of energy from large to small spatial scales, as expected for a sharpening front, with the key difference that ageostrophic advection enhances this transfer in the SG case.

In the depth-integrated budget (Fig.~\ref{fig:spec}a,c), buoyancy production serves as the source of kinetic energy and peaks at a wavenumber $k \sim 1$. Water sinks on the dense side of the front and rises on the light side of the front (Fig.~\ref{fig:config}). Some of the kinetic energy thus produced is transferred from large to small scales by geostrophic and ageostrophic (in the SG case) advection. The geostrophic advection term is negative at large scales and positive at small scales, indicating a downscale transfer. The ageostrophic advection term has a similar pattern but shifted to smaller scales and with a smaller amplitude. Some of the kinetic energy generated by buoyancy production and transferred to smaller scales by advection leads to a strengthening of the frontal jet, as exhibited by the positive tendency, which peaks at noticeably smaller scales than the buoyancy production. Much of the kinetic energy production, however, is balanced by the sink term.

The budgets at the surface (Fig.~\ref{fig:spec}b,d) mirror the depth-integrated budgets, except that pressure work takes the place of buoyancy production. At the surface, where $w = 0$, buoyancy production vanishes identically. Kinetic energy generated by buoyancy production in the interior is transported to the surface by pressure work.

\begin{figure}[t]
  \centering
  \includegraphics[width=\textwidth]{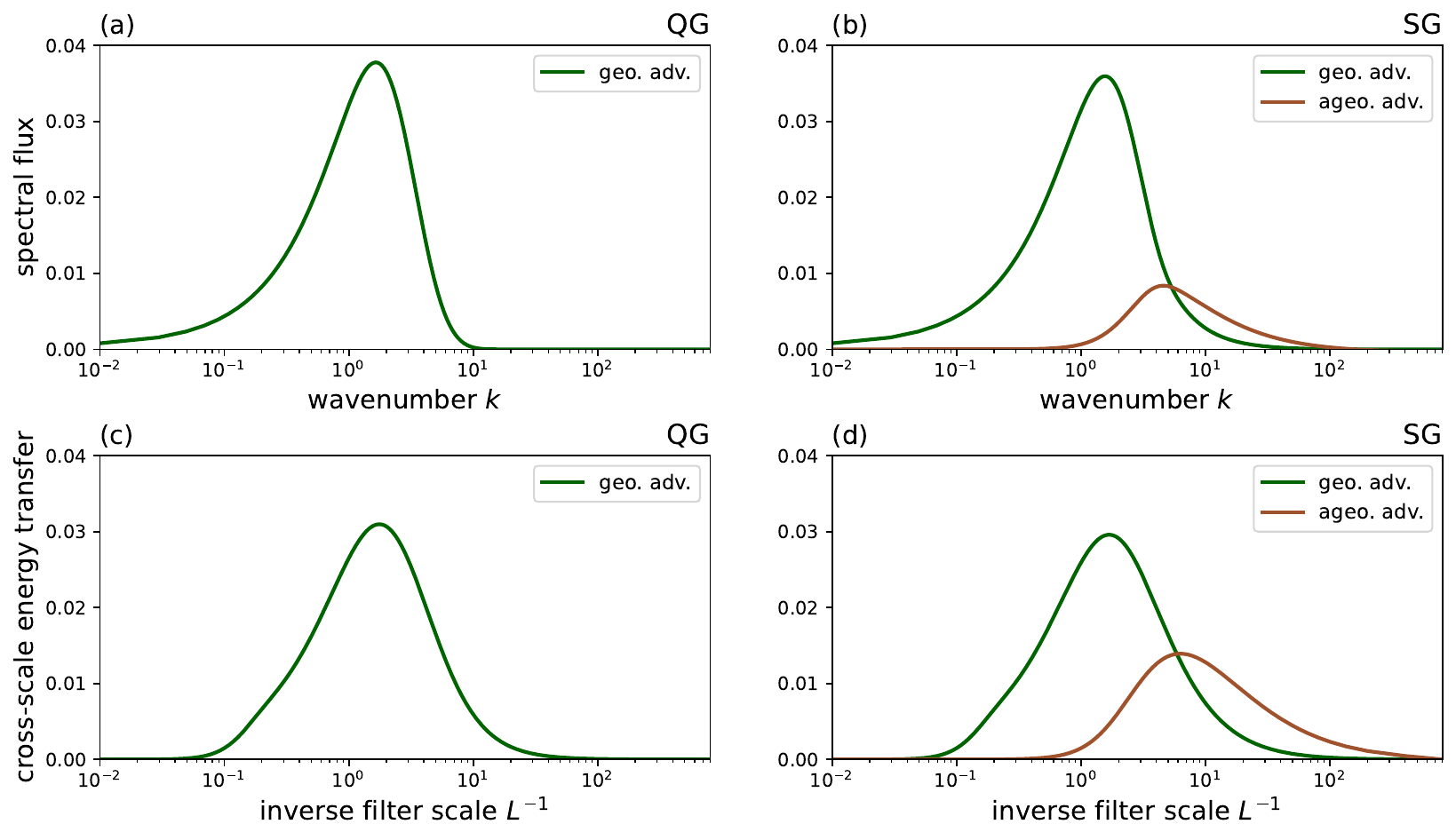}
  \caption{Horizontally integrated downscale kinetic energy transfer for the QG and SG fronts ($\gamma = 1$). Shown are the surface transfers calculated from the spectral budget (top) and the coarse-grained budget (bottom). For the SG case, the transfers are separated into geostrophic and ageostrophic contributions. The transfers are evaluated at $t = 1.33$, shortly before the collapse of the SG front. All variables are dimensionless (see Appendix~\ref{sec:nondim}).}
  \label{fig:spec_flux}
\end{figure}

To further illustrate the energy transfer to small scales and highlight the role of ageostrophic advection in the SG case, we calculate the spectral flux~\eqref{eq:spec_flux} of these two components at the surface (Fig.~\ref{fig:spec_flux}a,b). The geostrophic spectral flux is comparable between the QG and SG cases, with similar magnitudes and peaking at $k^{-1} = 0.61$ for QG and $k^{-1} = 0.64$ for SG. The ageostrophic spectral flux, present in SG only, peaks at a smaller scale ($k^{-1}=0.22$) and dominates the scale transfer at small scales. This behavior is entirely expected, given that it is ageostrophic advection that accelerates frontogenesis in SG. It should be kept in mind, however, that the ageostrophic surface circulation is horizontally convergent on the cyclonic side of the front and divergent on the anticyclonic side (Fig.~\ref{fig:config}). The convergence on the cyclonic side collapses the front and causes downscale energy transfer, but the divergence on the anticyclonic side weakens the buoyancy gradient and broadens the front there---an upscale transfer. The downscale transfer of kinetic energy in the spectral budget is an integral over both regions, indicating that the downscale transfer caused by convergence on the cyclonic side dominates. The spectral transfer may mask substantial compensation between the distinct behaviors on the two sides of the front. We take this as motivation to next pursue a localized budget using the coarse-graining framework.

\subsection{Coarse-grained kinetic energy budget}
\label{sec:coarse}

Although the spectral approach provides important information on the magnitude and sense of cross-scale energy transfer, it operates in wavenumber space and reflects spatially integrated behavior, giving no insight into the location of the energy transfer in physical space. To complement the spectral budget, we therefore apply the coarse-grained budgets \eqref{eq:qg_coarse} and~\eqref{eq:sg_coarse} to the sharpening front, focusing as before on $\gamma = 1$ and $t = 1.33$, immediately before frontal collapse in the SG case.

The Gaussian filter employed here for coarse graining offers an optimal compromise between resolution in spectral and physical space. While it is less sharp in spectral space than the Fourier filter used in the spectral budget, it yields a spatially integrated scale transfer~$\Pi$ that is qualitatively similar to the spectral flux at the surface (Fig.~\ref{fig:spec_flux}c,d). 
The geostrophic scale transfer peaks at a filter scale $L = \num{.57}$ in QG and $L = \num{.60}$ in SG, similar to the respective inverse wavenumbers in the spectral budget. 
The geostrophic scale transfers are somewhat reduced in magnitude, however, and their peaks are broadened. 
The ageostrophic scale transfer in SG peaks at $L = 0.16$, and its peak magnitude is elevated compared to the spectral budget, giving the impression of a relatively more important contribution from ageostrophic advection to the total scale transfer than in the spectral fluxes. Despite the quantitative differences, the qualitative behavior is similar between these two descriptions of scale transfer. We next demonstrate the additional spatial information that the coarse-graining approach offers.

\begin{figure}[t]
  \centering
  \includegraphics[width=\textwidth]{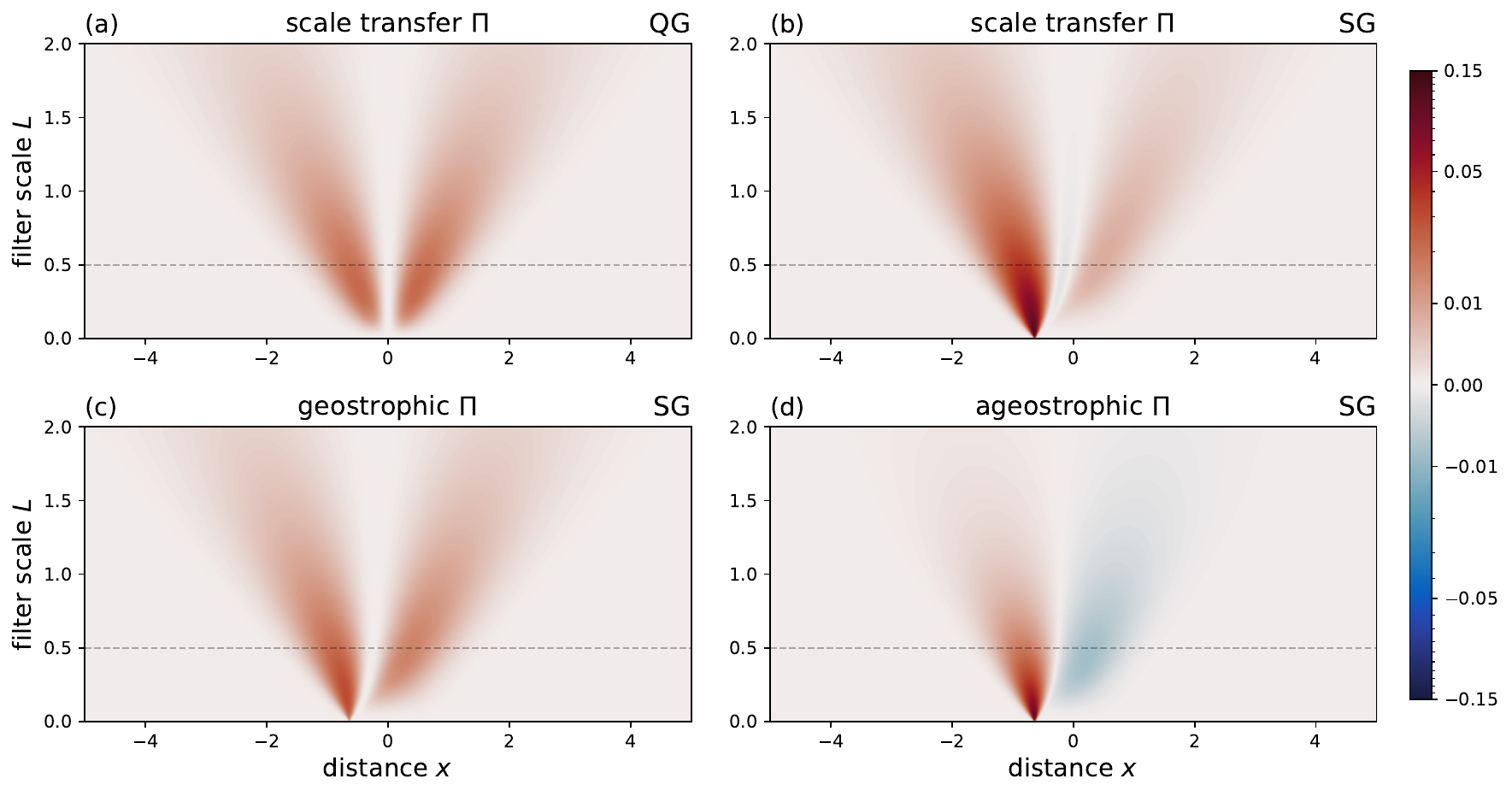}
  \caption{Surface cross-scale kinetic energy transfer~$\Pi$ from the coarse-grained budget. (a)~Geostrophic scale transfer for the QG front. (b)~Full scale transfer for the SG front ($\gamma = 1$). (c)~Geostrophic contribution to the scale transfer for the SG front. (d)~Ageostrophic contribution to the scale transfer for the SG front. The budgets are evaluated at $t = 1.33$, shortly before the collapse of the SG front. The dashed line at $L = 0.5$ shows the filter scale used in Fig.~\ref{fig:coarse}. Note the arsinh-scaled color map. All variables are dimensionless (see Appendix~\ref{sec:nondim}).}
  \label{fig:filter}
\end{figure}

In physical space, the scale transfer shows distinct maxima on either side of the surface front, increasingly concentrated near the front with decreasing filter scales (Fig.~\ref{fig:filter}a,b). This pattern is symmetric in QG because only the strain field sharpens the front through geostrophic advection. In SG, there is a marked asymmetry, with a much stronger downscale transfer on the cyclonic side of the front than on the anticyclonic side. This asymmetry arises because the front itself is asymmetric, so even geostrophic advection by the strain field produces an asymmetric pattern (Fig.~\ref{fig:filter}c). More importantly, the ageostrophic scale transfer has opposite signs on the two sides of the front (Fig.~\ref{fig:filter}d). Ageostrophic advection causes a downscale transfer on the cyclonic side and an upscale transfer on the anticyclonic side. This is expected because the ageostrophic circulation is horizontally convergent on the cyclonic side and divergent on the anticyclonic side (Fig.~\ref{fig:config}). On the anticyclonic side, while the downscale transfer caused by geostrophic advection dominates over the upscale transfer by ageostrophic advection, there is a substantial amount of cancellation, such that the overall downscale transfer is weakened substantially (Fig.~\ref{fig:filter}b--d). On the cyclonic side, the ageostrophic downscale transfer is comparable to the geostrophic downscale transfer, and it dominates over the geostrophic one at sufficiently small filter scales (Fig.~\ref{fig:filter}c,d).

\begin{figure}[p]
  \centering
  \includegraphics[width=0.85\textwidth]{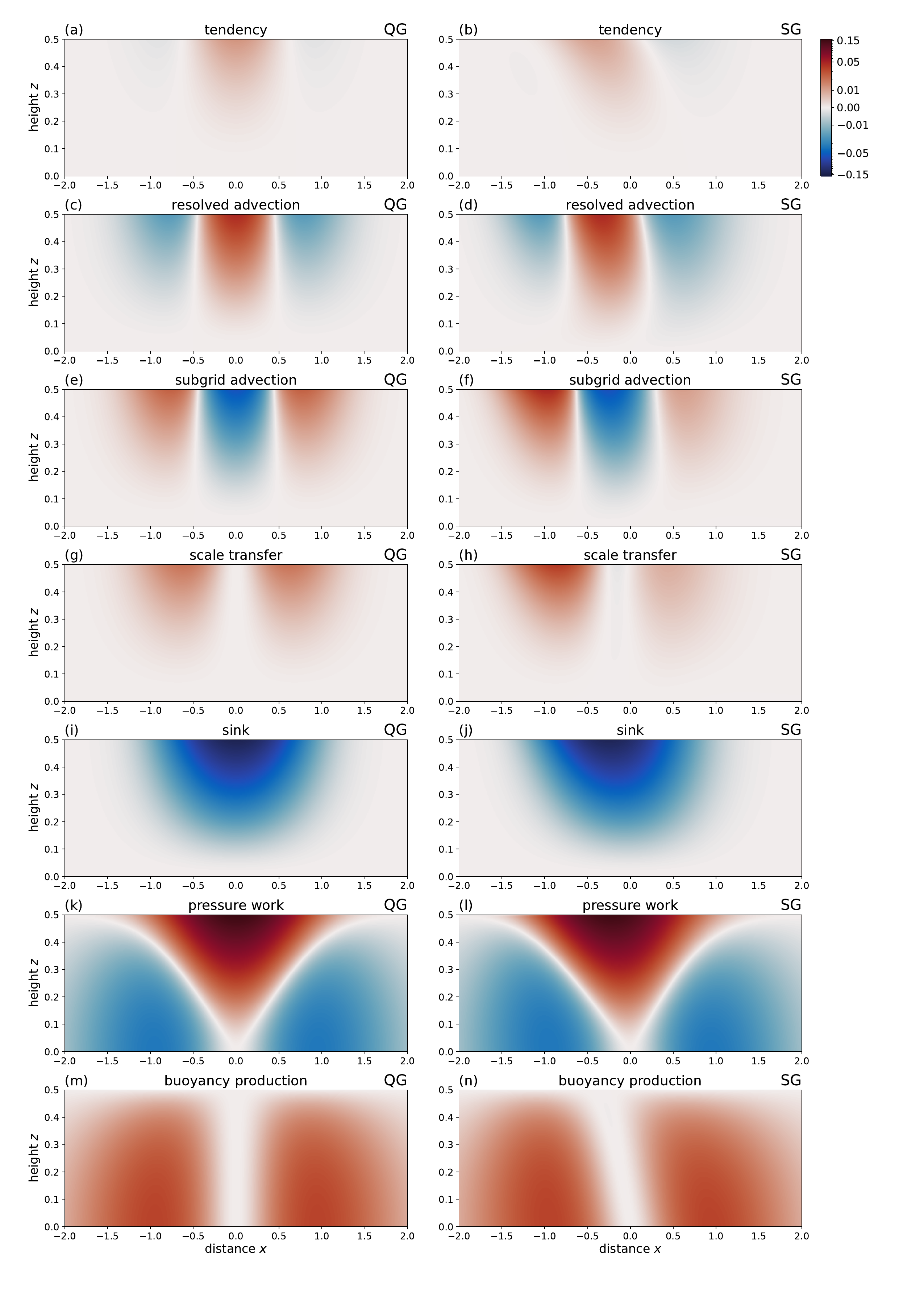}
  \caption{Coarse-grained kinetic energy budget for QG and SG fronts ($\gamma = 1$). The terms in \eqref{eq:qg_coarse} and \eqref{eq:sg_coarse} are shown with a sign convention as if all terms except the tendency (a,b) and cross-scale transfer (g,h) were on the right-hand sides of these equations. A positive scale transfer, $\Pi > 0$, corresponds to a downscale transfer and therefore a sink of filtered energy; for all other terms, the sign indicates whether the term is a source (positive) or sink (negative) of the filtered energy. The budgets are evaluated at $t = \num{1.33}$, with a spatial filter scale of $L = 0.5$.  Note the arsinh-scaled color map and that only the top half of the domain is shown. All variables are dimensionless (see Appendix~\ref{sec:nondim}).}
  \label{fig:coarse}
\end{figure}

The scale transfers and their spatial asymmetry in SG extend into the interior, as illustrated at the filter scale $L = 0.5$ (Fig.~\ref{fig:coarse}g,h). We note that, as exhibited in the spectral budget, the scale transfers are only part of the full kinetic energy budget, with a number of other terms exceeding their magnitudes (Fig.~\ref{fig:coarse}). In particular, buoyancy production from upwelling on the light side of the front and downwelling on the dense side is the net source of kinetic energy (Fig.~\ref{fig:coarse}m,n). Pressure work concentrates the kinetic energy into the front (Fig.~\ref{fig:coarse}k,l), and the sink term from stretching and shear production offsets this gain of kinetic energy in the front (Fig.~\ref{fig:coarse}i,j). Resolved advective transport moves kinetic energy from the flanks to the center of the front (Fig.~\ref{fig:coarse}c,d), but its effect is strongly compensated by the subgrid transport (Fig.~\ref{fig:coarse}e,f). The result is a net advective transport from the anticyclonic to the cyclonic side in the SG case, and a weak residual in the QG case. The scale transfers should be interpreted in this context of the full kinetic energy budget.

\section{Discussion}
\label{sec:discussion}

\begin{figure}[t]
  \centering
  \includegraphics[width=\textwidth]{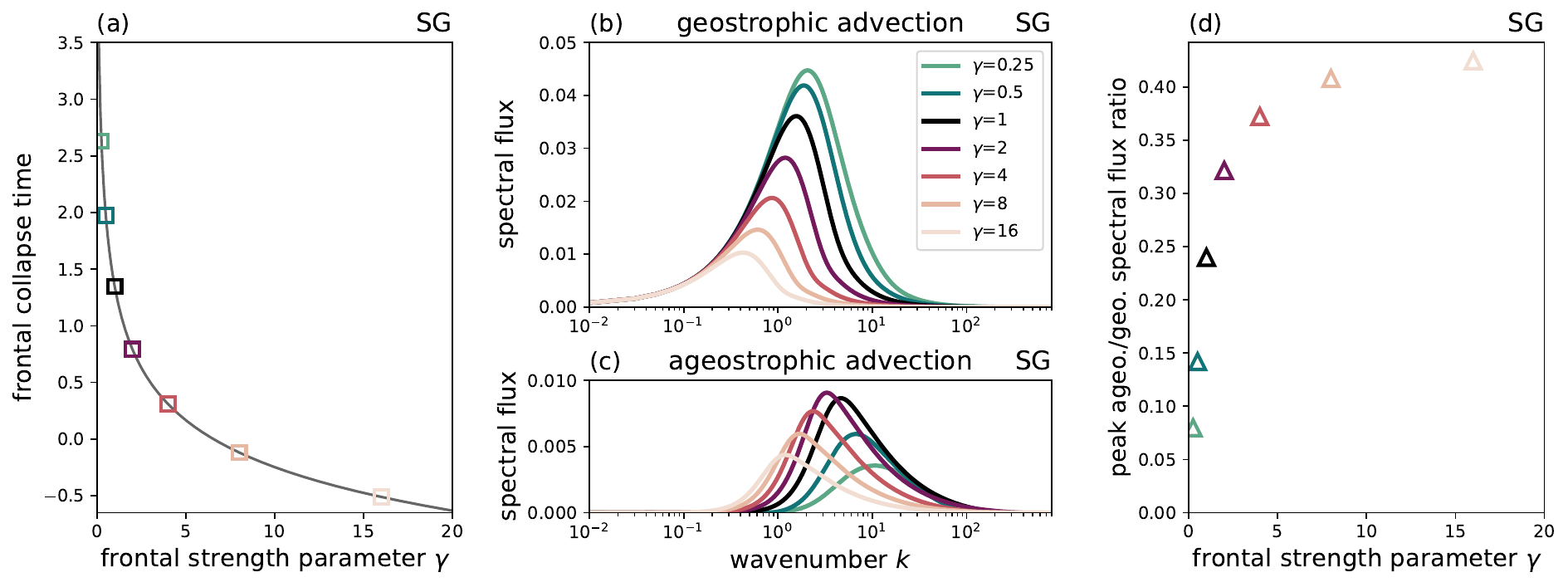}
  \caption{The impact of the frontal strength parameter~$\gamma$ on the SG frontal evolution. (a)~Frontal collapse time as a function of $\gamma$. Colored boxes indicate $\gamma$ values selected for comparison of energy transfer in (b,c). The reference case $\gamma = 1$, analyzed in detail in previous figures, is marked in black. (b)~Surface spectral energy flux from geostrophic advection for selected values of $\gamma$ shortly before the respective frontal collapse time. (c)~Same as (b) but for ageostrophic advection. (d)~The ratio of peak ageostrophic to geostrophic spectral fluxes for selected values of $\gamma$. All variables are dimensionless (see Appendix~\ref{sec:nondim}).}
  \label{fig:gamma}
\end{figure}

\subsection{Dependence on frontal strength}
\label{sec:gamma}

While QG frontogenesis is independent of it, the evolution of an SG front depends on the frontal strength parameter~$\gamma = \Delta / N^2 H$, the ratio between the buoyancy drop across the front and the vertical buoyancy contrast.
So far, we have considered the $\gamma = 1$ case, but the surface ocean has fronts with varying strength. In winter, in particular, one might expect strong lateral buoyancy gradients, weak stratification, and deep mixed layers, so $\gamma > 1$. According to the observations compiled by \textcite{whalen2025}, a typical front in the wintertime mixed layer may have $\Delta = \qty{e-3}{\meter\per\second\squared}$, $N^2 = \qty{4e-6}{\per\second\squared}$, and $H = \qty{100}{\meter}$, which gives $\gamma = 2.5$. We therefore consider briefly how $\gamma$ affects the scale transfer of energy in SG frontogenesis.

In the nondimensionalization of the SG equations, $\gamma$ emerges naturally (see Section~\ref{sec:nondim_short} and Appendix~\ref{sec:nondim}). While the SG equations in geostrophic coordinates are independent of~$\gamma$, the coordinate transformation from geostrophic to physical coordinates does depend on it. The larger~$\gamma$, the greater the distortion of the front and the faster the collapse. The collapse time as a function of~$\gamma$ can be calculated by setting the Jacobian~$J$ of the transformation to zero, which provides the time at which the coordinate transformation ceases to be single-valued (Fig.~\ref{fig:gamma}a). We diagnose the spectral energy transfer at \num{.01} dimensionless time units before the collapse time.

An increase in the frontal strength~$\gamma$ leads to a higher relative contribution from ageostrophic cross-scale energy fluxes to the total energy transfer (Fig.~\ref{fig:gamma}b--d).
As $\gamma$ increases from 0.25 to 16, the peak spectral flux due to geostrophic advection decreases monotonically. At $\gamma = 16$, the peak is reduced by 77\% compared to the $\gamma = 0.25$ case. This decrease can be understood using the QG solution evaluated at the varying SG collapse times. The earlier collapse time for increasing $\gamma$ mean that the QG front---and the SG front in geostrophic coordinates---is broader and weaker at the time of diagnosis, explaining the decrease of the peak flux and shift to larger scales. 
In contrast, the peak spectral flux due to ageostrophic advection increases with $\gamma$ up to $\gamma = 2$, after which it decreases again.
A detailed interpretation of this non-monotonic behavior is beyond the scope of this study.
The ratio of the peak spectral fluxes due to ageostrophic and geostrophic advection increases monotonically from 8\% at $\gamma = 0.25$ to 42\% at $\gamma = 16$, highlighting the growing relative importance of ageostrophic dynamics in strong fronts.
This ratio asymptotes to 44\%, the value the SG solution achieves for a front in a zero-PV layer, which corresponds to the $\gamma \to \infty$ limit (see below). 
The qualitative nature of the spectral energy budget across this range of~$\gamma$ is similar to the $\gamma = 1$ case analyzed at length above (Section~\ref{sec:energy}).

\subsection{Comparison with primitive-equation simulations}

The spatial structure of kinetic energy scale transfer in QG and SG frontogenesis diverges in some aspects from that in realistic primitive-equation simulations \parencite{srinivasan2023}.
The simple models of frontogenesis agree with the diagnosis of downscale energy transfer at fronts in primitive-equation simulations. As the front gets sharpened by the strain field as well as the ageostrophic secondary circulation, the frontal jet narrows, and energy is transferred to small scales (Fig.~\ref{fig:spec}--\ref{fig:coarse}). 
In SG frontogenesis, the ageostrophic circulation by itself produces a dipole of energy transfer: a downscale transfer on the cyclonic side, where the ageostrophic circulation is horizontally convergent, and an upscale transfer on the anticyclonic side, where the ageostrophic circulation is horizontally divergent (Fig.~\ref{fig:filter}d).
This ageostrophic transfer is consistent with what is found in realistic primitive-equation simulations \parencite[Fig.~\ref{fig:non};][]{srinivasan2023}.
However, realistic simulations exhibit a relatively stronger contribution from the ageostrophic circulation to the overall scale transfer. \textcite{srinivasan2023} found the ageostrophic transfer to dominate over geostrophic transfers, such that the ageostrophic upscale transfer overpowered geostrophic downscale transfer on the anticyclonic side of the front, leaving a dipole in the overall transfer and a tight correlation with the surface divergence (cf., Fig.~\ref{fig:non}).

We emphasize that our setup considers an isolated front, whereas realistic simulations produce fronts embedded in a turbulent mesoscale eddy field. While our setup helps isolate the processes causing energy transfer in a simple context, the energy exchange more broadly should involve mesoscale and submesoscale eddies. The sink term in our frontal setup should be thought of as a stand-in for the interaction with the surrounding eddy field, which has confluent as well as diffluent regions. The sink term would be part of the scale transfer term for a spatially confined front \parencite[cf.,][]{shakespeare_generation_2015}, which may explain part of the differences with the primitive-equation simulation diagnostics. Perhaps more importantly, primitive-equation simulations include the additional physics of advection of ageostrophic momentum, and they parameterize mixed-layer turbulence, which affect the evolution of fronts and the strength of the ageostrophic circulation. We now discuss in turn these two possible explanations for the differences with realistic primitive-equation simulations.

\subsection{Advection of ageostrophic momentum}
\label{sec:general-front}

One possible explanation for the enhanced role of the ageostrophic flow and the pronounced dipole of downscale and upscale transfer in realistic primitive-equation simulations is that the advection of ageostrophic momentum, which is neglected in SG dynamics, becomes important \parencite{shakespeare_generation_2015,barkan19}.
We therefore consider briefly how the advection of ageostrophic momentum modifies the energetics of a front in the special case of a zero-PV layer.
We focus on this case because ageostrophic advection is most important in SG solutions in the $\gamma \to \infty$ limit (Fig.~\ref{fig:gamma}d) and because the zero-PV case is most relevant to the ocean's mixed layer.
The theory of \textcite{shakespeare_generalized_2013} makes a milder approximation than SG theory and thereby takes some of the additional effects of the advection of ageostrophic momentum into account.
For the zero-PV case, the \textcite{shakespeare_generalized_2013} solution for strain-driven frontogenesis can be separated into a balanced solution and two near-inertial wave modes.
The balanced part of the solution reverts to the SG solution, suggesting that the SG solution accurately describes the evolution of the front even when the strain is strong ($\alpha \sim f$).
The approximations made by \textcite{shakespeare_generalized_2013}, however, break down once the front is sufficiently sharp, which is the regime that we are interested in here.
Thus, we instead assess the impact of the advection of ageostrophic momentum in the full primitive equations.

For the case of a zero-PV layer, we modify the nondimensionalization used in the rest of the paper by setting $N^2 = \Delta H$, because the stratification away from the front now vanishes and $N$ is no longer available as an independent scale. The nondimensional primitive equations are then
\begin{align}
  \varepsilon^2 \left( \DD{u}{t} - u \right) - v &= -\pp{\phi}{x}, \label{eqn:pe-umom} \\
  \DD{v}{t} + v + u &= 0, \label{eqn:pe-vmom} \\
  0 &= -\pp{\phi}{z} + b, \\
  \pp{u}{x} + \pp{w}{z} &= 0, \\
  \DD{b}{t} &= 0, \label{eqn:pe-buoy}
\end{align}
where the material derivative is
\begin{equation}
  \DD{}{t} = \pp{}{t} + (-x + u) \pp{}{x} + w \pp{}{z}.
\end{equation}
The only dimensionless parameter in this problem is $\varepsilon \equiv \alpha/f$, the Rossby number of the strain field. SG theory assumes $\varepsilon \ll 1$ and replaces~\eqref{eqn:pe-umom} with geostrophic balance.
Keeping all terms here, the kinetic energy budget is
\begin{multline}
  \pp{}{t} \frac{\varepsilon^2 u^2 + v^2}{2} + \pp{}{x} \left[ (-x + u) \frac{\varepsilon^2 u^2 + v^2}{2} + u \phi \right] \\
  + \pp{}{z} \left[ w \frac{\varepsilon^2 u^2 + v^2}{2} + w \phi \right] + \frac{3}{2} v^2 - \frac{1}{2} \varepsilon^2 u^2 = wb.
\end{multline}
Compared to the SG energy budget, the kinetic energy now includes a cross-front component~$\frac{1}{2} \varepsilon^2 u^2$ that is not necessarily negligible.
There is now geostrophic (strain field) and ageostrophic advection of this cross-front kinetic energy, which produces additional scale transfer terms in the spectral and coarse-grained budgets analogous to the terms arising from the advection of along-front momentum.
The cross-front advection of the strain field also produces an additional source of kinetic energy~$\frac{1}{2} \varepsilon^2 u^2$ that is similar in nature to the sink term~$-\frac{3}{2} v^2$ \parencite{shakespeare_generation_2015}.

To solve the primitive equations \eqref{eqn:pe-umom} to~\eqref{eqn:pe-buoy}, we apply the coordinate transformation
\begin{equation}
  X = e^{t} x, \quad Z = z, \quad T = t, \label{eq:new_coord}
\end{equation}
which transforms the material derivative into
\begin{equation}
  \DD{}{t} = \pp{}{T} + e^T u \pp{}{X} + w \pp{}{Z}.
\end{equation}
Note the difference in the coordinate transformation from that applied in the SG case.
The coordinates used here zoom in on the sharpening front and thereby make the advection by the strain field implicit, but they leave the ageostrophic cross-front advection in the transformed equations. In the transformed coordinates, the primitive equations become
\begin{align}
  \varepsilon^2 \left( \pp{}{T} + e^T u \pp{}{X} + w \pp{}{Z} - 1 \right) u - v &= -e^T \pp{\phi}{X}, \\
  \left( \pp{}{T} + e^T u \pp{}{X} + w \pp{}{Z} + 1 \right) v + u &= 0, \\
  0 &= -\pp{\phi}{Z} + b, \\
  e^T \pp{u}{X} + \pp{w}{Z} &= 0, \\
  \left( \pp{}{T} + e^T u \pp{}{X} + w \pp{}{Z} \right) b &= 0.
\end{align}
We solve these equations on a staggered grid ($u$~on $X$~faces, $w$~on $Z$~faces) with an implicit treatment of the pressure terms, which together with the continuity equation forms an algebraic saddle-point problem \parencite[e.g.,][]{strang_computational_2007}.
We include no explicit diffusion and solve with a grid spacing of \num{.025} in $X$ and $Z$, and we apply solid walls at $X = \pm 10$, far away from the front.
We initialize $u$, $v$, and $b$ with the QG solution at $T = -2$ and step forward with a time step of \num{e-3}.
The solutions are not sensitive to these numerical choices.

\begin{figure}[tp]
  \centering
  \includegraphics[width=\textwidth]{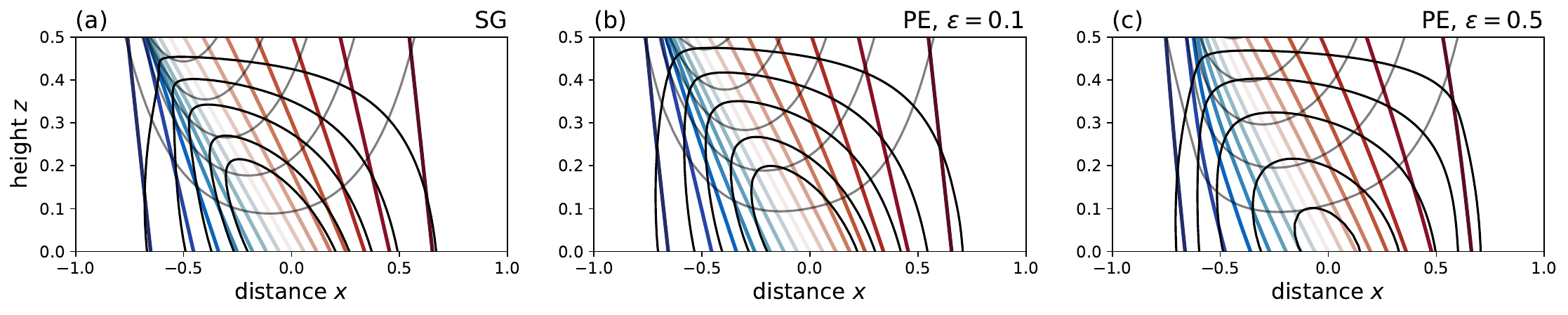}
  \caption{Fronts in a zero-PV layer as described by semi-geostrophic theory and the full primitive equations. Shown are (a)~the SG solution, (b)~the primitive-equation solution with $\varepsilon = \alpha/f = 0.1$, and (c)~the primitive-equation solution with $\varepsilon = 0.5$ that is outside of the SG regime. The colored contours show the buoyancy field, the black contours the stream function of the cross-frontal circulation, and the gray contours the along-front flow. All three solutions are shown at $t = 0.81$. Note that only the top half of the domain is shown. All variables are dimensionless (see Section~\ref{sec:general-front}).}
  \label{fig:zeropvfronts}
\end{figure}

We compare the cases $\varepsilon = 0.1$ and $0.5$ to the corresponding SG solution ($\varepsilon = 0$) that we obtain as before but with the modified nondimensionalization appropriate for the zero-PV case. We evaluate all solutions at $T = 0.81$, which is shortly before the collapse of the SG front (Fig.~\ref{fig:zeropvfronts}). All three solutions show a cross-frontal circulation that is confined to the frontal zone and shows a marked asymmetry between broad upwelling on the anticyclonic side of the front and narrow downwelling on the cyclonic side of the front. The SG solution is very similar to the primitive-equation solution with $\varepsilon = 0.1$, as expected. As the primitive-equation solution is pushed out of the SG regime at $\varepsilon = 0.5$, the upwelling on the anticyclonic side also narrows. The magnitude of the cross-frontal circulation is similar between all three cases.

\begin{figure}[tp]
  \centering
  \includegraphics[width=\textwidth]{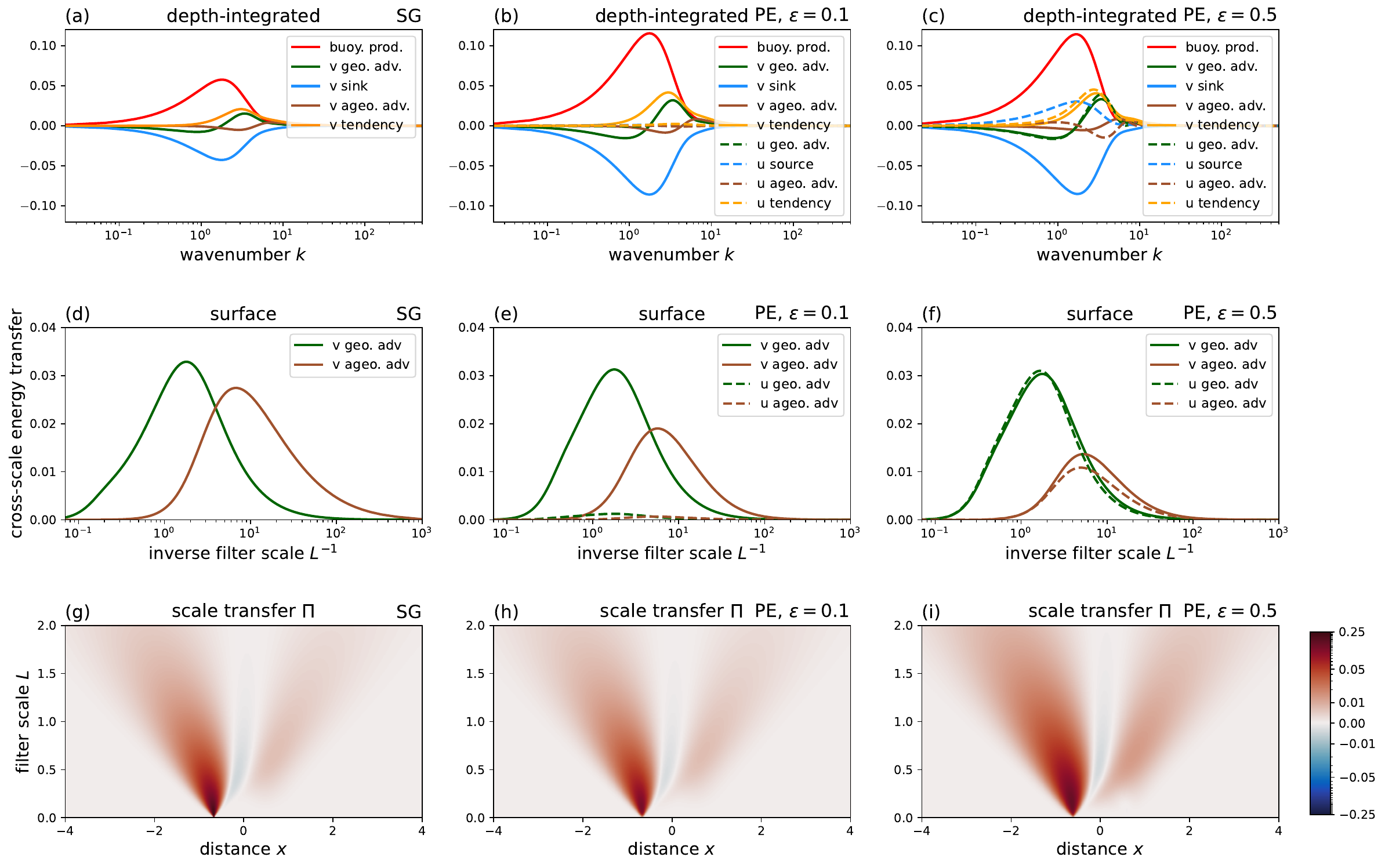}
  \caption{Spectral kinetic energy budget and cross-scale energy transfer for the semi-geostrophic (left column) and primitive-equation solutions ($\varepsilon = 0.1$, middle column; $\varepsilon = 0.5$, right column) to strain-driven frontogenesis in a zero-PV layer. The spectral budget terms are shown with a sign convention as if all terms except the tendency were on the right-hand sides of the kinetic energy budget. The budgets are evaluated at $t = \num{0.81}$, shortly before the collapse of the SG front. (a--c)~Spectral kinetic energy budget integrated over the depth of the domain. (d--f)~Horizontally integrated scale transfers~$\Pi$ from coarse-graining at the surface ($z = \frac{1}{2}$). (g--i)~Surface scale transfers~$\Pi$ as a function of cross-front position and filter scale. All variables are dimensionless (see Section~\ref{sec:general-front}).}
  \label{fig:zeropvbudgets}
\end{figure}

We now assess the depth-integrated spectral kinetic energy budgets of these solutions (Fig.~\ref{fig:zeropvbudgets}a--c). We restrict ourselves to the depth-integrated budgets because the primitive equations now allow for vertical advection of kinetic energy, such that the advection terms can be a source or sink of energy at a given level---they can be considered redistributive between scales only upon vertical integration. The spectral energy budgets of this SG solution and the primitive-equation solution with $\varepsilon = 0.1$ are similar to those of the SG solutions presented previously, with buoyancy production being the source of kinetic energy, geostrophic and ageostrophic scale transfer moving kinetic energy downscale, the strain interaction providing a substantial sink, and the acceleration of the along-front flow peaking at around the same scale as where the geostrophic scale transfers deposit energy. In the primitive-equation solution with $\varepsilon = 0.1$, the additional terms in the budget remain small. At $\varepsilon = 0.5$, these terms now contribute substantially. While all terms from the SG budget are similar in shape and magnitude, there now is a substantial source from the interaction with the strain, substantial geostrophic and ageostrophic scale transfer due to advection of cross-front momentum, and a substantial contribution from the tendency of the cross-front component of the kinetic energy.

At the surface, the scale transfers as diagnosed using the coarse-graining framework show a substantial contribution to the downscale flux from the ageostrophic circulation and a substantial role of advection of cross-front momentum in the $\varepsilon = 0.5$ solution (Fig.~\ref{fig:zeropvbudgets}d--f). In all three solutions, the ageostrophic advection dominates the downscale transfer at small scales, although its peak remains smaller than that of the geostrophic downscale transfer. The advection of cross-front momentum remains negligible at $\varepsilon = 0.1$. At $\varepsilon = 0.5$, it produces scale transfers very similar to those arising from the advection of along-front momentum, both geostrophic and ageostrophic. It is interesting to note that the ageostrophic scale transfer is upscale in the depth integral (Fig.~\ref{fig:zeropvbudgets}c), in contrast to the downscale transfer at the surface.

The full scale transfers from the coarse-graining analysis as a function of cross-front position and filter scale still show downscale transfer on both sides of the front (Fig.~\ref{fig:zeropvbudgets}g--i). These scale transfers have a remarkably similar structure between the three zero-PV solutions. The $\varepsilon = 0.5$ case has scale transfers with a slightly greater magnitude, but the pattern and relative magnitudes are very similar to those of the SG solution. This pattern of the scale transfers remains in stark contrast to that seen in realistic primitive-equation simulations, suggesting that the neglect of the advection of cross-front momentum in SG theory cannot by itself explain this difference, at least for the special case of a zero-PV layer and the parameter regime explored here.

Whether this conclusion also holds for a non-zero PV layer remains unclear. In that case, the generalized theory of \textcite{shakespeare_generalized_2013,shakespeare_spontaneous_2014,shakespeare_spontaneous_2015} predicts spontaneous wave emission and confinement of the front, which could provide an additional pathway for scale transfer beyond that captured by the balanced SG theory. While such emission at fronts does not appear to feature prominently in realistic primitive-equation simulations \parencite{barkan_eddyinternal_2024,delpech_eddyinternal_2024}, the energy transfers in this regime warrant further investigation in future work. The picture is significantly complicated, however, because the lack of time scale separation between the frontal evolution and waves leaves the concept of a balanced part of the circulation ill-defined.

\subsection{Mixed layer turbulence}
\label{sec:turbulence}

A key idealization in our setup is that we study adiabatic and frictionless frontogenesis, neglecting the effects of mixed-layer turbulence. Like the sharpening of the front by the strain field, mixing of momentum and buoyancy throws the front out of thermal-wind balance and thus induces a secondary circulation that acts to restore it. 
The net effect of mixing can be either frontogenetic or frontolytic, depending on the relative strengths of momentum and buoyancy mixing, the stages of frontal development, background mechanical forcing, as well as on whether horizontal or vertical mixing dominates \parencite{garrett1981,thompson2000ekman,mcwilliams2015,crowe2018,bodner2020,dauhajre2025}.
For example, frictional forcing from vertical mixing can drive frontal slumping that accelerates surface sharpening \parencite{thompson2000ekman}, while the combination of vertical mixing and shear may spread and weaken surface fronts after the initial frontogenetic phase \parencite{crowe2018}.
If the parameterized mixing is frontogenetic, a strengthened ageostrophic circulation will enhance the dipole of energy transfer due the convergence on the cyclonic side and divergence on the anticyclonic side of the front, and the resulting ageostrophic scale transfers can overpower the geostrophic transfer that dominates in the inviscid and adiabatic case.
\textcite{barkan19} identified this as the dominant mechanism driving frontogenesis in their realistic simulations, and they diagnosed a similar role of convergence in the sharpening of fronts from drifter data. \textcite{torres_airborne_2024} also found a tell-tale correlation between convergence and downscale energy transfer in airborne observations of submesoscale fronts, suggesting that this effect is possibly at play in the real ocean as well. 
Studying the impact of turbulent mixing on the energetics of frontogenesis in a controlled setup---and a careful comparison with the impact of the advection of ageostrophic momentum (Section~\ref{sec:general-front})---would help strengthen this interpretation of the simulated and observed phenomenology.

\section{Conclusion}

Frontogenesis as described by simple balanced theories produces a downscale transfer of kinetic energy (Fig.~\ref{fig:spec_flux}). A sharpening front is associated with a narrowing along-front jet and therefore a transfer of kinetic energy to small scales. A geostrophic strain field by itself produces such downscale transfer of kinetic energy, as described by QG theory \parencite{stone1966}. Ageostrophic advection, included in the SG theory of frontogenesis \parencite{hoskins72}, tilts over the front and accelerates frontogenesis. On the cyclonic side, the resulting horizontal convergence accelerates the sharpening of the buoyancy gradient and enhances the downscale transfer of kinetic energy (Fig.~\ref{fig:filter}d). On the anticyclonic side, horizontal divergence slows the sharpening of the buoyancy gradient and is associated with an upscale transfer of kinetic energy (Fig.~\ref{fig:filter}d). In the net, ageostrophic advection enhances the kinetic energy transfer to small scales because the downscale transfer on the cyclonic side is stronger than the upscale transfer on the anticyclonic side (Fig.~\ref{fig:filter}b,d).

The adiabatic and frictionless frontogenesis considered here displays a smaller ageostrophic contribution to the scale transfer of kinetic energy than realistic simulations and observations suggest (Fig.~\ref{fig:non}). In the idealized front, the dipole in the ageostrophic transfer---downscale on the cyclonic side of the front, upscale on the anticyclonic side---causes a distinct asymmetry in the overall transfer, but it is not strong enough to overpower the universally downscale geostrophic contribution to give a dipole in the overall transfer. This remains true over a wide range of frontal strengths (Fig.~\ref{fig:gamma}) and in primitive-equation solutions for a zero-PV layer (Fig.~\ref{fig:zeropvbudgets}). We suspect that an enhancement of the ageostrophic circulation by non-conservative processes could flip this balance \parencite{shakespeare_spontaneous_2014,shakespeare_generation_2015}, but this is left for future work.

While the submesoscale scale transfer of kinetic energy has received heightened attention for its suspected importance as a route to dissipation \parencite[e.g.,][]{capet2008-3,muller_routes_2011,schubert_submesoscale_2020,naveira_garabato_kinetic_2022,srinivasan2023}, it is important to interpret this transfer in the context of the full kinetic energy budget. In the idealized theories of frontogenesis considered here, buoyancy production plays a major role in the energetics (Figs.~\ref{fig:spec},~\ref{fig:coarse}). The ageostrophic circulation lifts up light water and pushes down dense water as it slumps the front (Fig.~\ref{fig:config}). Pressure work transports the resulting kinetic energy into the surface front. The scale transfer moves some of this kinetic energy to smaller scales, playing an important role in the sharpening of the frontal jet. Equally important, however, is the sink of kinetic energy density due to the along-front stretching and shear production associated with the strain field. It would be interesting to extend this scale-resolved analysis of the full kinetic energy budget to more realistic simulations, in which the fronts are embedded in a turbulent eddy field.

\section*{Acknowledgments}

We thank Andrew Thompson for insightful discussions and the two reviewers for their constructive and thoughtful feedback, which has strengthened the manuscript. YB acknowledges support from the National Aeronautics and Space Administration (NASA) Science Mission Directorate FINESST program under Award No.~80NSSC21K1635. JC acknowledges support from NASA under Award Nos.~80NSSC20K1140 and 80NSSC24K1652.

\section*{Data availability statement}
Example scripts of QG and SG strain-driven frontal development and kinetic energy budget calculations in both physical and spectral space are available at: \url{https://doi.org/10.5281/zenodo.17102819}.

\appendix

\section{Nondimensionalization}
\label{sec:nondim}

Following the scaling stated in Section~\ref{sec:nondim_short}, we present the full nondimensionalized QG and SG kinetic energy equations, which serve as the basis for the spectral and coarse-grained kinetic energy budget results in Section~\ref{sec:energy}.

For the QG case, we follow standard procedure. Using the scales~\eqref{eqn:nondimqg} in the QG kinetic energy equation~\eqref{eq:qg_ke}, every term in the budget acquires a scale $\alpha \Delta^2/N^2$. The dimensionless form of this equation thus becomes independent of any parameters:
\begin{equation}
  \pp{}{\tilde t} \frac{ \breve v_0^2}{2} + \pp{}{\breve x} \bigg( - \breve x \frac{\breve v_0^2}{2} + \breve u_1 \breve \phi_0 \bigg) + \pp{}{\breve z} \bigg( \breve w_1 \breve \phi_0 \bigg) + \frac{3}{2} \breve v_0^2 = \breve w_1 \breve b_0,
  \label{eq:qg_ke_non}
\end{equation}
where $\breve \cdot$ marks the nondimensionalized quantities. Note that we dropped this decoration in the main text starting in Section~\ref{sec:energy}.

The SG solution in geostrophic coordinates is identical to the QG solution in geostrophic coordinates and therefore also remains independent of any parameters. The coordinate transformation back to physical space, however, introduces the dimensionless parameter $\gamma = \Delta / N^2 H$ that characterizes the frontal strength.
The nondimensionalized coordinate transformation is
\begin{equation}
    \breve x = \breve X - \gamma \breve v_g,
\end{equation}
and its Jacobian is
\begin{equation}    
    \breve J = \bigg( 1 - \gamma\pp{\breve v_g}{\breve X} \bigg)^{-1},
\end{equation}
so the amount of distortion of the solution depends on the choice of~$\gamma$.
The modified geopotential is
\begin{equation}
   \breve \phi = \breve \Phi - \gamma \frac{\breve v_g^2}{2},
\end{equation}
and the horizontal ageostrophic velocity becomes
\begin{equation}
    \breve{u} = \breve{\tilde{u}} - \gamma \breve w \pp{\breve v_g}{\breve z}.
\end{equation}
The parameter~$\gamma$ also appears in the nondimensionalized kinetic energy budget:
\begin{equation}
  \pp{}{\breve t} \frac{\breve v_g^2}{2} + \pp{}{\breve x} \bigg[ (- \breve x + \gamma \breve u) \frac{\breve v_g^2}{2} + \breve u \breve \phi \bigg] + \pp{}{\breve z} \bigg[ \gamma \breve w \frac{\breve v_g^2}{2} + \breve w \breve \phi \bigg] + \frac{3}{2} \breve v_g^2 = \breve w \breve b.
  \label{eq:sg_ke_non}
\end{equation}
As in the QG case, all terms share a dimensional scale $\alpha \Delta^2/N^2$, except for the ageostrophic advection of momentum by $\breve u$ and~$\breve w$, which carry an additional factor of~$\gamma$. 

The nondimensionalized spectral and coarse-grained kinetic energy budgets are derived and diagnosed based on \eqref{eq:qg_ke_non} and~\eqref{eq:sg_ke_non}, following the steps laid out in Sections \ref{sec:spec_eqn} and~\ref{sec:coarse_eqn}. 
In the spectral analysis, the wavenumber~$k$ is scaled with $f/NH$, the inverse of the characteristic length scale. 
In coarse-graining, the Gaussian filter scale~$L$ is the length scale at which the scale energy transfer is assessed, and is therefore nondimensionalized by $NH/f$, sharing the same scaling as the cross-front coordinate~$x$.

\end{document}